\documentclass[review]{elsarticle}

\usepackage{amsmath}
\usepackage{amsfonts}
\usepackage{gensymb}
\usepackage[dvipsnames]{xcolor}
\usepackage{subcaption}
\usepackage{xcolor}
\usepackage[utf8x]{inputenc}
\usepackage{chngcntr}
\counterwithout{figure}{subsection}

\begin{document}

\begin{frontmatter}

\title{Light Output Quenching in Response to Deuterium-ions and Alpha Particles and Pulse Shape Discrimination in Deuterated Trans-stilbene}

\author[mymainaddress]{J. Zhou\corref{mycorrespondingauthor}}
\cortext[mycorrespondingauthor]{Corresponding author.}
\ead{jianxin6@illinois.edu}
\author[mymainaddress]{N. Gaughan}
\author[mysecondaryaddress]{F. D. Becchetti}
\author[mysecondaryaddress]{R. O. Torres-Isea}
\author[mythirdaddress]{M. Febbraro}
\author[myfourthaddress]{N. Zaitseva}
\author[mymainaddress]{A. Di Fulvio}

\address[mymainaddress]{Department of Nuclear, Plasma, and Radiological Engineering, \\University of Illinois at Urbana-Champaign,
                        \\ Urbana, IL, USA}
\address[mysecondaryaddress]{Department of Physics, University of Michigan, Ann Arbor, MI, USA}
\address[mythirdaddress]{Oak Ridge National Laboratory, Oak Ridge, TN, USA}
\address[myfourthaddress]{Lawrence Livermore National Laboratory, Livermore, CA, USA}

\begin{abstract}
We characterized the light output response of a new 140 cm$^3$ stilbene-d$_{12}$ crystal up to 14.1 MeV neutron energies using a coincidence neutron scattering system. We also characterized its light output response to alpha particles in the 5 to 6~MeV energy range.
The excellent PSD capability of the stilbene-d$_{12}$ detector allowed us to select light pulses produced by particles of increasing ionization density, namely electrons, protons, deuterium-ions, and alpha particles. The measured fast decay component of the light pulses is increasingly quenched as the ionization density of the particle in the crystal increases.
Consistently with this finding, the Birks' quenching parameter of alpha particles is approximately 8.5 times larger compared to the quenching of deuterium ions, produced by neutron scattering interactions. The reported experimental characterization will allow high-fidelity modeling of the detector enabling its application for fast-neutron detection and spectroscopy in nuclear physics, radiation protection, nuclear security, and non-proliferation.  

\end{abstract}


\end{frontmatter}


\section{Introduction} \label{s:intro}

Deuterated scintillators detect neutrons via the deuteron recoils produced by n(d,d')n' scattering reactions. The fore-back asymmetry of n(d,d')n' scattering produces distinct peaks in the light-output response to monoenergetic neutrons. Because of this feature, complex neutron spectra can be reconstructed effectively from the measured light-output response through the unfolding technique~\cite{Zhu2019} without the need of a time-of-flight measurement setup. Therefore, deuterated scintillators outperform $^1$H-based scintillators in fast-neutron spectroscopy~\cite{Brooks1979, Lawrence2013, Gaughan2021}. 

Several deuterated liquid organic scintillation detectors have been developed, such as deuterated-benzene$~$\cite{Ojaruega2011} and deuterated-xylene~\cite{Becchetti2016,DiFulvio2017}, and they are used in many applications that demand the knowledge of the neutron field, ranging from nuclear physics \cite{Borella2007, Febbraro2013,Febbraro2014} to nuclear safeguards~\cite{DiFulvio2017, Egner2021}. Accurate knowledge of the light-output response to neutrons is needed to unfold the interacting neutron spectrum from the measured light-output spectrum~\cite{Gaughan2021}.
In this paper, we characterized for the first time the neutron light output and light quenching parameters of a 140~cm$^3$ deuterated trans-stilbene crystal (stilbene-d$_{12}$). The crystal was grown at Lawrence Livermore National Laboratory (LLNL) in 2019 using the solution growth method from deuterated styrene precursor~\cite{Carman2018} . \par
In our previous work~\cite{Gaughan2021}, we characterized the light output response of a smaller 32~cm$^3$ stilbene-d$_{12}$ crystal to quasi-monoenergetic neutrons up to 4.4~MeV. In this work, we used a larger (140~cm$^3$) crystal and designed a scattering-based experimental setup based on a D-T neutron source to extend the light output response up to 14.1~MeV neutron energy. 
We also measured the stilbene-d$_{12}$ light output response to alpha particles emitted by $^{237}$Np, $^{239}$Pu, $^{241}$Am, and $^{244}$Cm. This analysis allowed us to compare the stilbene-d$_{12}$ light quenching in response to the interaction with deuterons, in the case of D-T neutrons, and with alpha particles, in the case of the actinides.
A quenched light-output response to particles with different ionization densities determines the pulse shape discrimination (PSD) capability exhibited by organic scintillation detectors. PSD enables the discrimination between different interacting radiation based on the specific amplitudes and time constants of the fluorescence signal following ionization caused by ions with different linear energy transfer (LET). 
In this work, we further characterized the ionization quenching parameters of the fluorescence signal in stilbene-d$_{12}$ in response to the interaction with electrons, protons, deuterium ions, and alpha particles. \par

\section{Materials and methods} \label{s:method}

We designed a coincidence neutron scattering experimental setup based on a D-T neutron source to measure the neutron light output response of stilbene-d$_{12}$ up to 14.1~MeV. We also used mono-energetic alpha sources to characterize the light output response to 5-6~MeV alpha particles. 

\subsection{Scintillation light and ionization quenching}
\label{sec:Meth_decay_time}

\begin{figure}[htbp!]
    \centering
    \includegraphics[width=0.4\textwidth]{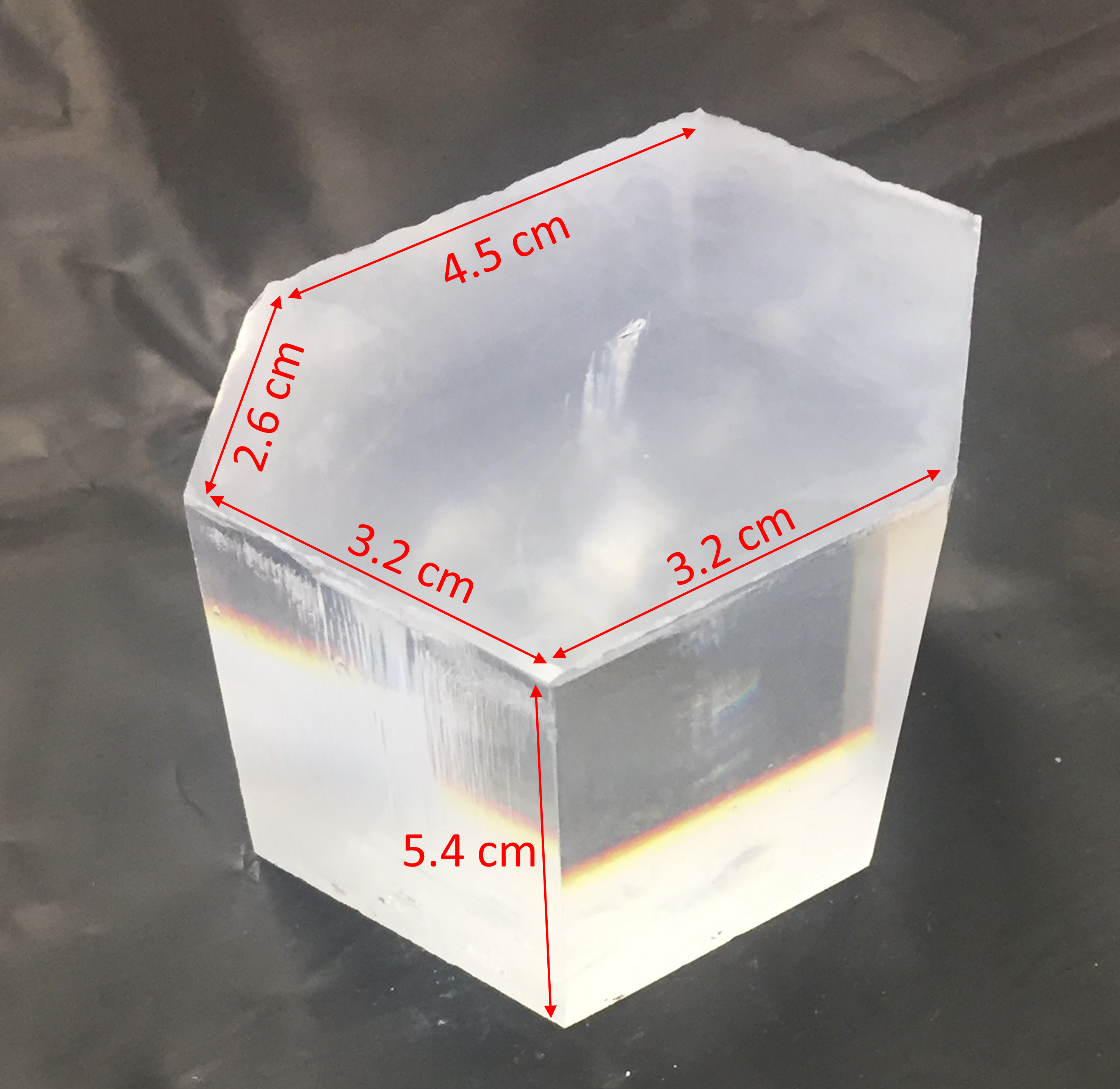}
    \caption{Stilbene-d$_{12}$ crystal, non-equilateral hexagonal prismatic shape.}
    \label{f:dsb_crystal}
\end{figure}

Figure \ref{f:dsb_crystal} shows the 140 cm$^3$ stilbene-d$_{12}$ crystal that we used in this work. 
Upon interaction with ionizing radiation, the detectable luminescence mainly comes from the $\pi$ electrons, which exist in $\pi$ bond (a type of covalent chemical bond) in organic scintillators. The ionization energies first excite $\pi$ electrons to the singlet states (S$_1$, S$_2$, S$_3$, ...), and then the de-excitation of the electrons emits the detectable luminescence. Because of the fast time of this de-excitation process (tens of nanoseconds), the emitted light is referred to as the ``fast component" of the fluorescence signal. Besides the singlet states, the triplet states (T$_1$, T$_2$, T$_3$, ...) also exist during ionization. The triplet states can annihilate and produce S$_1$ electrons. The annihilation rate increases with ionization density. These S$_1$ electrons will keep emitting luminescence in a longer time period due to the lifetime of T$_1$ states (hundreds of nanoseconds). Therefore, the emitted light through this process is referred to as the ``slow component" of the fluorescence signal.
The ionization quenching is the electron non-radiative de-excitation process, which results in an overall loss of scintillation light. It affects the pulse shape by changing the ratio of fast and slow components. In this paper, we measured the light pulses in the stilbene-d$_{12}$ in response to the interaction with electrons, protons, deuterium-ions, and alpha particles and quantified the time constants and intensities of fast and slow components and the quenching parameters.\par
We obtained light pulses of deuterium-ions, protons, and electrons using a D-T generator. The D-T neutrons can produce deuterium-ion pulses via n(d,d')n' scattering in the stilbene-d$_{12}$. They can also produce proton pulses via n(d,2n)p break-up reactions, which has a 3.35~MeV neutron energy threshold \cite{Brown2018}. The generator also emits gamma rays from neutron interactions with the generator tube, which can produce Compton electrons in the crystal. 
We measured alpha light pulses from four alpha sources, $^{237}$Np, $^{239}$Pu, $^{241}$Am, and $^{244}$Cm. Each source has a single alpha decay branch and emits monoenergetic alpha particles. The PSD technique was applied to classify the pulse types of the measured data. We first calculated the ``pulse total integral" by integrating each scintillation pulse over 370~ns, starting 4 ns before the time stamp of the pulse maximum. Then, we obtained the ``pulse tail integral" in a similar way (starting the integral 22 ns after the pulse maximum). Finally, we calculated a PSD parameter for each pulse, as the ratio between the ``pulse tail integral" and the ``pulse total integral" (TTR). We used the TTR to distinguish particle types. In each scatter-density plot where the TTR is reported as a function of the pulse integral, the discrimination line has the following functional form: $TTR=a \times x^2+b \times x+c$, where a, b, and c are constants and x is the pulse total integral. The discrimination lines divide pulses into different groups and each group represents a particle type.  We optimized the values of a, b, and c choosing the values that yielded the minimum number of misclassificated pulses.
For each pulse type, we normalized thousands of well-classified signals in selected light output ranges and averaged them to acquire the pulse templates. Electron, proton, and deuterium-ion pulses in 2-6 MeVee range were selected and alpha pulses in 0.25-0.5 MeVee range. Pulses within these ranges can be well-discriminated by their TTR values, as shown in Figure \ref{f:DT_Alpha_PSD}.
We used Equation~\ref{eqn: light_exp_decay} to describe the time profile of the pulse templates and characterized the light amplitudes and time constants of the fast and slow components.\par
\begin{equation}
    L(t)=-a_re^{(\frac{-t}{\tau_r})}+\frac{q_1}{\tau_1}e^{(\frac{-t}{\tau_1})}+\frac{q_2}{\tau_2}e^{(\frac{-t}{\tau_2})}+\frac{q_3}{\tau_3}e^{(\frac{-t}{\tau_3})}
    \label{eqn: light_exp_decay}
\end{equation}
The first term in Equation \ref{eqn: light_exp_decay} describes the rising edge of the light profile.,
$\tau_r$ is the rising edge time constant. The last three terms describe the light decay. The fast component comes from the direct S$_1$ de-excitation process and the light intensity decays exponentially. Therefore, we used one exponential term to describe it. $\tau_1$ represents the decay constant of the fast component and $q_1$ represents its weight. The slow component is due to the T$_1$ electrons, which are mainly produced from two processes (``inter-system crossing" and ``ion recombination"\cite{Birks1965}). Hence, we used two exponential terms to describe its time profile. $\tau_2$ and $\tau_3$ represent the decay constants of the slow components and $q_2$ and $q_3$ represent their weights.\par
The quenching coefficients of various particles can be quantitatively evaluated through the Birks' Equation~\ref{eqn:Birks_model}~\cite{Birks1965}. This model is commonly used to describe the relationship between the energy deposited and their light emissions ($L$). $dE/dr$ is the particle  stopping power in the scintillation materials, $A$ is the scintillation  efficiency, and $B$ is the quenching parameter, which demonstrates the quenching levels of various particles.

We used the methods described in the following sections to measure neutron and alpha light output responses. 
By fitting the responses with Equation~\ref{eqn:Birks_model}, we quantitatively compared the quenching levels of deuterium-ions and alpha particles through the parameter $B$. The stopping power in Equation~\ref{eqn:Birks_model} in stilbene-d$_{12}$ was calculated using the SRIM software~\cite{Ziegler2010}

\begin{equation}
    L(E)=\int_0^{E'} \frac{AdE}{1+BdE/dr}
    \label{eqn:Birks_model}
\end{equation}

\subsection{Neutron scattering kinematic}
\label{sec:Meth_scatter_kinematic}

Neutron detection in stilbene-d$_{12}$ mainly relies on the n(d,d’)n’ elastic scattering. If the neutron scattering angle ($\theta$) and the energy of the interacting neutron $E_n$ are known, Equations~\ref{eqn: scattering_kine1} and \ref{eqn: scattering_kine2} determine the energies of the scattered neutrons ($E'_n$) and the recoil nucleus ($E_r$), respectively. $A$ is the mass number of the target nucleus.\par

\begin{equation}
    E'_n=E_n\times [\frac{cos\theta+\sqrt{A^2-sin^2\theta}}{A+1}]^2
    \label{eqn: scattering_kine1}
\end{equation}

\begin{equation}
    E_r=E_n\times (1-[\frac{cos\theta+\sqrt{A^2-sin^2\theta}}{A+1}]^2)
    \label{eqn: scattering_kine2}
\end{equation}

\begin{figure}[htb!]
    \centering
    \includegraphics[width=0.5\textwidth]{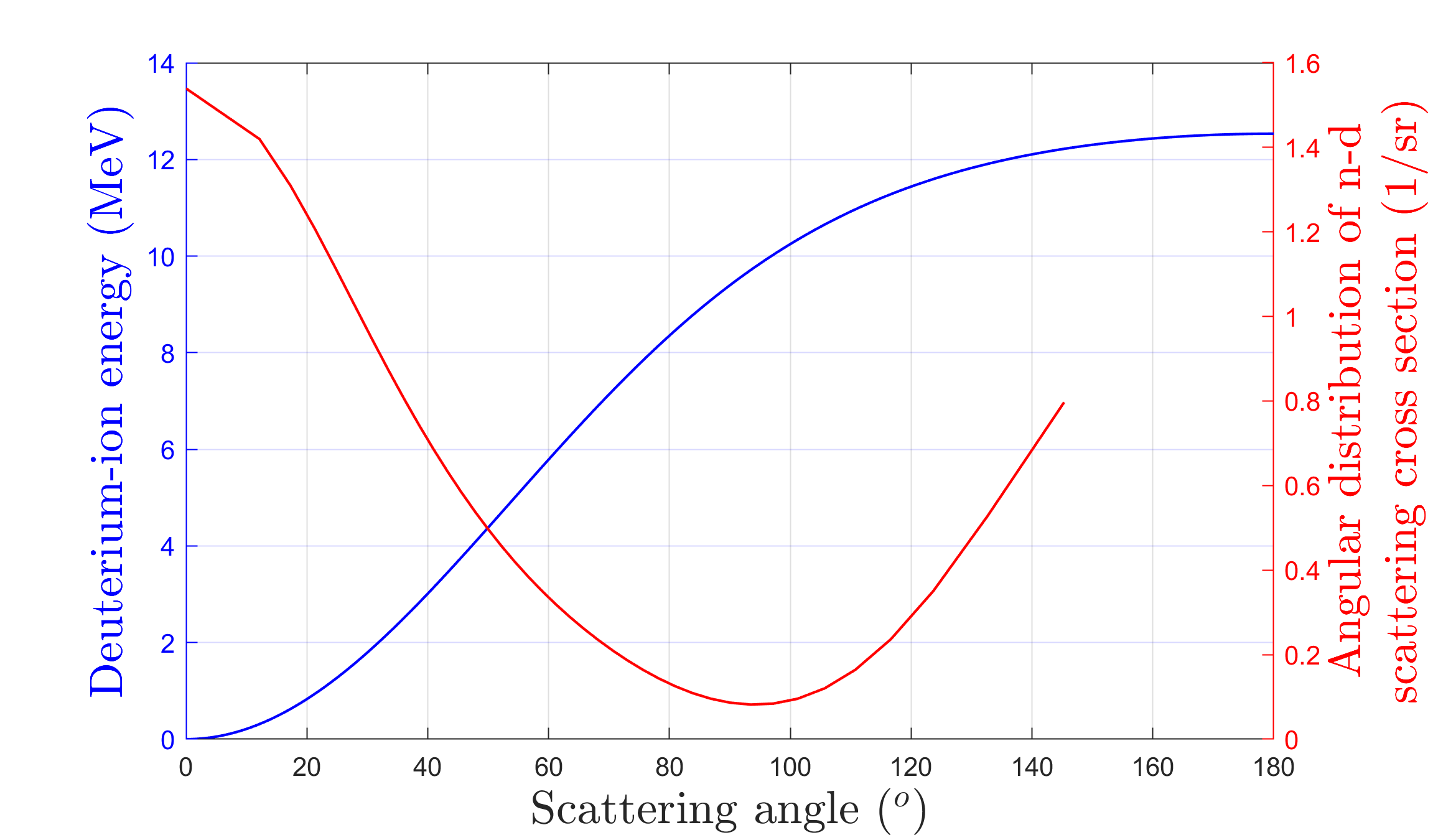}
    \caption{Energy transferred to the deuterium ion through scattering as a function to the scattering angles in the laboratory reference frame and angle-differential scattering cross section.}
    \label{f:scattering_energy_CrossSection}
\end{figure}

We used a D-T generator to characterize the neutron light output response. Through Equation~\ref{eqn: scattering_kine2}, we calculated the energies transferred to deuterium-ions with respect to various neutron scattering angles $\theta$, as shown in Figure~\ref{f:scattering_energy_CrossSection}.
The energy trasferred to the deuterium-ion increases with the neutron scattering angle.
 Figure~\ref{f:scattering_energy_CrossSection} also shows the non-isotropic angular distribution of the n-d scattering based on the ENDF/B-VIII.0 library~\cite{Brown2018}. 

\subsection{Experimental setup for light output response measurements}

This section describes the experimental methods that we used to characterize the neutron and alpha light output responses. We coupled the stilbene-d$_{12}$ crystal with a Hamamatsu H6559 photomultiplier assembly powered by a CAEN DT5533EN HV power supply. The detector signals were collected by a CAEN DT5730 desktop digitizer, which transmitted the digitizer pulses to a desktop computer via USB. We used a 10~dB, 1-GHz bandwidth, signal attenuator to match the PMT output with the digitizer 0-2~V amplitude input range. The use of the attenuator allowed to detect light pulses in the 0.05-12.7 MeVee light output range. The signals were processed by Python and Matlab custom software. We measured the pulse integral spectrum of a 1~$\mu$Ci $^{137}$Cs source and 1~$\mu$Ci $^{22}$Na source to calibrate the detector response in terms of electron-equivalent (eVee) units. We calibrated the detector response to electrons using the full energies deposited by Compton scattered electrons corresponding to the 85$\%$ of the Compton edges produced by 511~keV and 662~keV gamma rays from $^{22}$Na and $^{137}$Cs sources corresponding to 341~keVee and 478~keV electron energy deposited, respectively \cite{Becchetti2018}. The obtained calibration curve was: light output (MeVee) = 0.6727 (MeVee/(V$\times$number of samples))$\times$Pulse integral (V$\times$number of samples). The acquisition threshold of our detectors was approximately 50 keVee, corresponding to 450 keV and 400 keV neutron energy deposited in the EJ-309 and the stilbene-d12, respectively. We used the same acquisition and calibration procedures for the neutron and alpha measurements.
 
\subsubsection{Neutron light output response characterization with a coincidence neutron scattering system}
\begin{figure}[!htbp]
		\centering
		\subfloat[]{\includegraphics[width=0.5\textwidth]{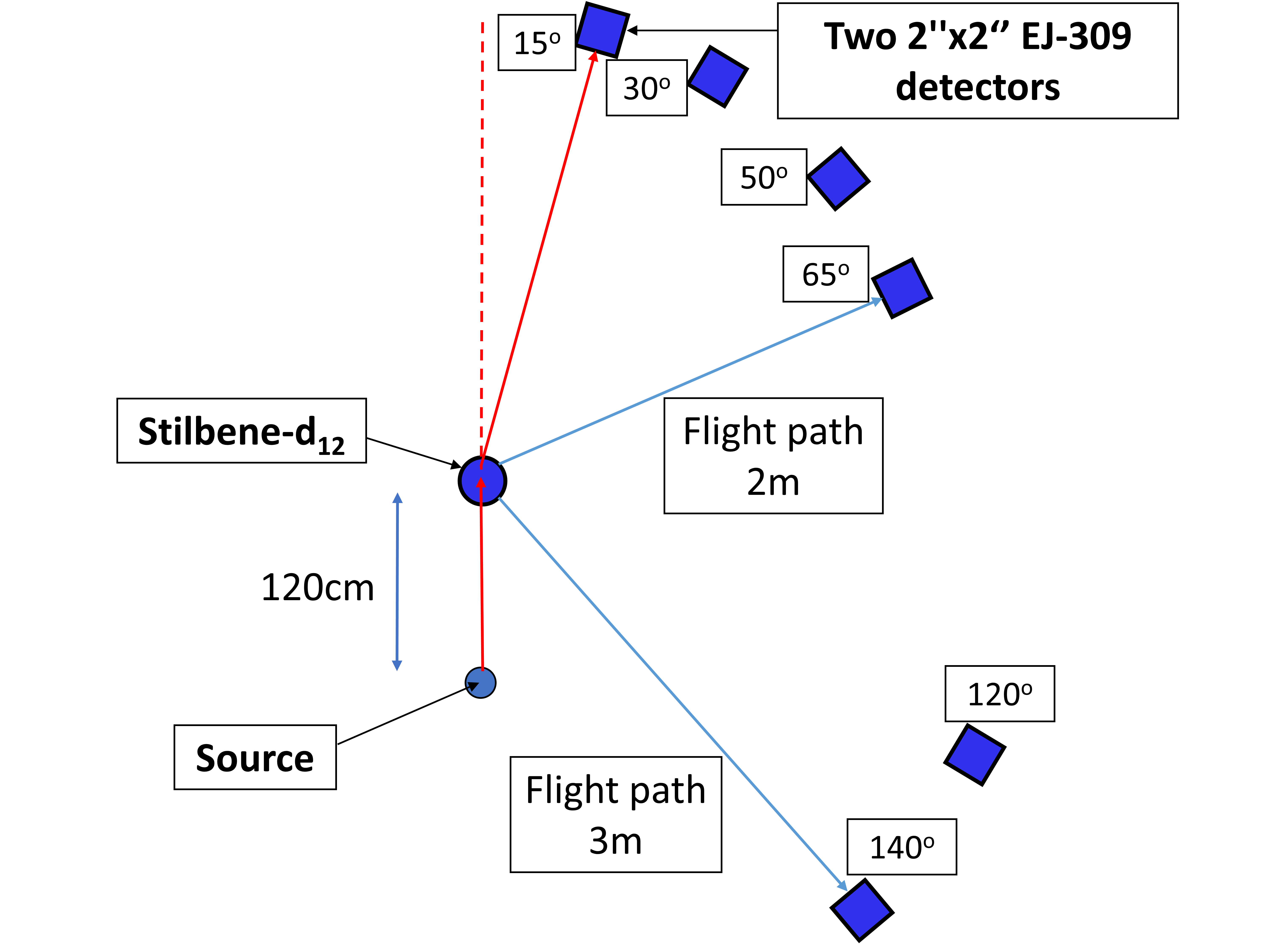}}
		\subfloat[]{\includegraphics[width=0.5\textwidth]{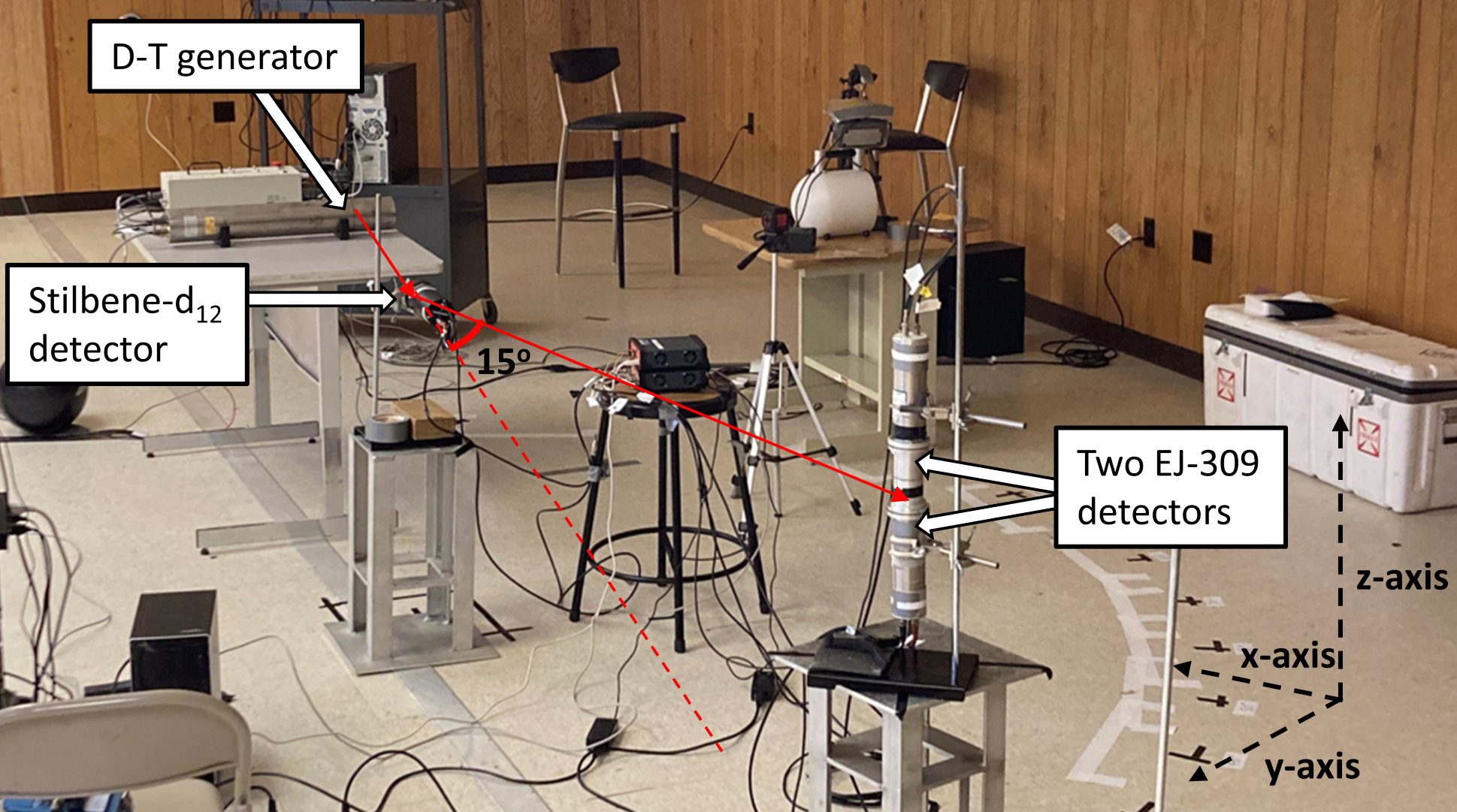}}
		\caption{(a) The schematic of the n-d scattering experiment for light output measurements. (b) The experimental setup for the 15 degree scattering angle measurement.}
		\label{f:scatter_setup}
	\end{figure}

We designed a coincidence neutron scattering system to extend the neutron light output response, based on the n(d,d’)n’ scattering information in Section~\ref{sec:Meth_scatter_kinematic}. Figure~\ref{f:scatter_setup} shows the experimental setup at the Nuclear Measurement Laboratory at UIUC. We placed the stilbene-d$_{12}$ detector 120~cm away from the D-T generator and used it as a scattering medium. We placed two EJ-309 detectors (5.08 cm diameter by 5.08 cm length) at various locations to measure the scattered neutrons. The two EJ-309 detectors were aligned along the Z-axis, as shown in Figure~\ref{f:scatter_setup} (b), to maximize the coincidence rate while minimizing the angular uncertainty in the X-Y plane. We detected pulses in the detectors within a 150~ns coincidence window. Since the primary neutron energy and neutron scattering angles were known, we calculated the energies of the scattered deuterium ions and neutrons using Equations~\ref{eqn: scattering_kine1} and ~\ref{eqn: scattering_kine2}, respectively. We determined the theoretical flight time from the stilbene\-d$_{12}$ to the EJ-309 detectors. Through the flight time, we further discriminated the true scattering events from the accidental coincidences within the short time window and measured the stilbene-d$_{12}$ light output values corresponding to energies deposited by the scattered deuterium ions. We performed the PSD technique on both stilbene-d$_{12}$ and EJ-309 detectors to reject random coincidence events triggered by gamma rays. The PSD rejection method is described in Section~\ref{sec:Meth_decay_time}. The time-of-flight measurements require the precise measurements of the neutron flight time. In theory, the events occurring at the same time would be detected at an averaged time difference of 0 ns. However, the asymmetries in the channel acquisition stages, such as different cable lengths, could determine a bias in the timing difference. Therefore, we calibrated the timing difference between channels using the 511~keV gamma rays emitted by the $^{22}$Na positron annihilation. 
	
We performed measurements at six scattering angles (15\degree, 30\degree, 50\degree, 65\degree, 120\degree, and 140\degree). Angles at which the scattering differential cross sections are relatively large were selected. The distance between the stilbene-d$_{12}$ and the EJ-309 detectors were 2~m for measurements at 15\degree, 30\degree, 50\degree, and 65\degree. For the back-scattering measurements (120\degree and 140\degree), we increased the distance to 3~m to mitigate the random coincidence events at backward angles due to the fact that in this case the EJ-309 detectors were closer to the neutron source. We operated the D-T generator at the 45~$\mu$A current and the 60~kV voltage. These settings yielded and source strength of ~6$\times$10$^7$ neutrons per second in the 4$\pi$ solid angle. The measurement time for each angle varied from 7 hours to 21 hours.\par
We performed the neutron measurements at the Nuclear Measurement Laboratory at the University of Illinois at Urbana-Champaign. The facility has 12.7-cm-thick walls made of wood and aluminum and it is surrounded by an open space (46 meter x 46 meter). This design minimizes neutron thermalization and back-scattering \cite{Daniel2020}. Moreover, the sensitivity of the applied coincidence neutron scattering system is inherently minimized to the neutrons that are slowed down or back-scattered. As described in Figure \ref{f:scatter_setup}(a), the coincidence windows of the measurements were based on the theoretical flight time of neutrons from the stilbene-d1$_{2}$ to EJ-309 detectors. Therefore, lower energy neutrons that would take longer to reach the EJ-309 detectors were rejected.

\subsubsection{Characterization of light output response to alpha particles}


We used three alpha disk sources to characterize the light output in response to alpha particles in the 5-6 MeV energy range. The activity, energy, and type of alpha sources are reported in Table~\ref{t:alpha_source_detail}. 
\begin{table}[hbt!]
    \centering
    \caption{Details about the measured alpha sources. }
    \begin{tabular}{ |c|c|c|c|c| }
    \hline 
  Source ID & Radionuclides & $\alpha$ energy & Activity (10/01/2021) \\
   & & (MeV) & ($\mu$Ci)\\
    \hline 
  $\#$1 & $^{241}$Am &  5.637 & 0.994\\
  $\#$2 & $^{237}$Np, $^{241}$Am, $^{244}$Cm & 4.959, 5.637, 5.902 & 1.5\\
  $\#$3 & $^{239}$Pu, $^{241}$Am, $^{244}$Cm &  5.244, 5.637, 5.902 & 1.5\\
    \hline 
    \end{tabular}
    \label{t:alpha_source_detail}
\end{table}

Alpha particles emitted by the measured sources have a short range of approximately 50 $\mu$m in plastic; therefore, they cannot penetrate the detector plastic-based light-tight coating, encompassing diffuse reflector PTFE tape and light-tight black tape, which is approximately 1~mm thick. Therefore, the measurements were performed in a dark chamber with the alpha sources placed directly on the detector surface. The detector surface not exposed to alpha particles was wrapped in PTFE and black tape.
\par
We determined the light output in response to the monoenergetic alpha particles emitted by the sources listed in Table~\ref{t:alpha_source_detail}.
However, for the two mixed source disks, it was not possible to discriminate the single peaks corresponding to the alpha sources.
Therefore, we used up to three Gaussian distributions to fit the responses to the two mixed source disks and obtain the light output values corresponding to the energy deposited by each nuclide.

\section{Results} \label{s:results}
In this section, we show the PSD capability of the 140~cm$^3$ stilbene-d$_{12}$ detector and the time constants corresponding to the fast and slow fluorescence components in response to electrons, protons, deuterium-ions, and alpha particles. We also present the light output responses to neutrons and alpha particles and their ionization quenching parameters.

\subsection{PSD and scintillation time constant}

\begin{figure}[!htb]
		\centering
		\subfloat[]{\includegraphics[width=0.5\textwidth]{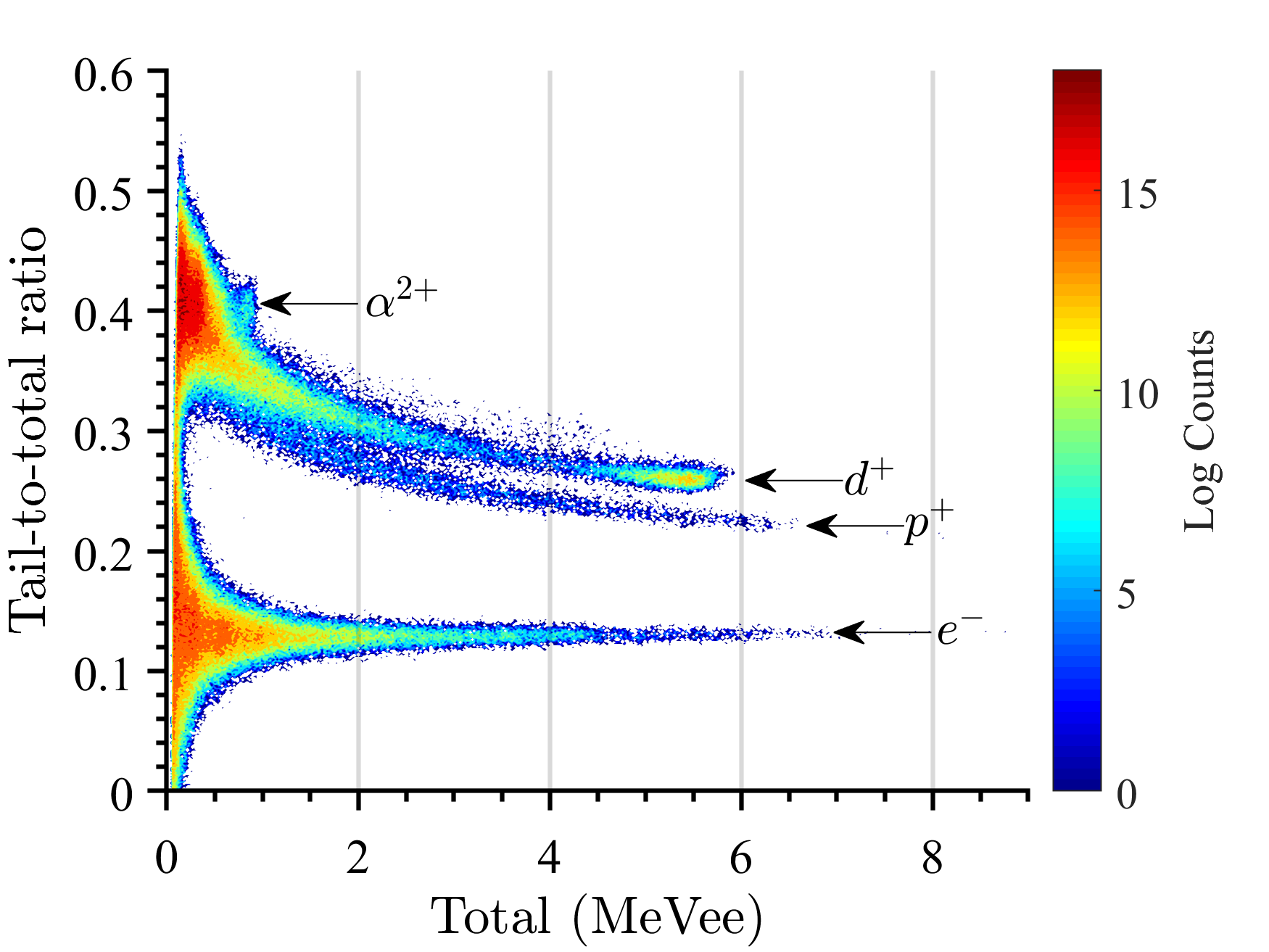}}
		\subfloat[]{\includegraphics[width=0.5\textwidth]{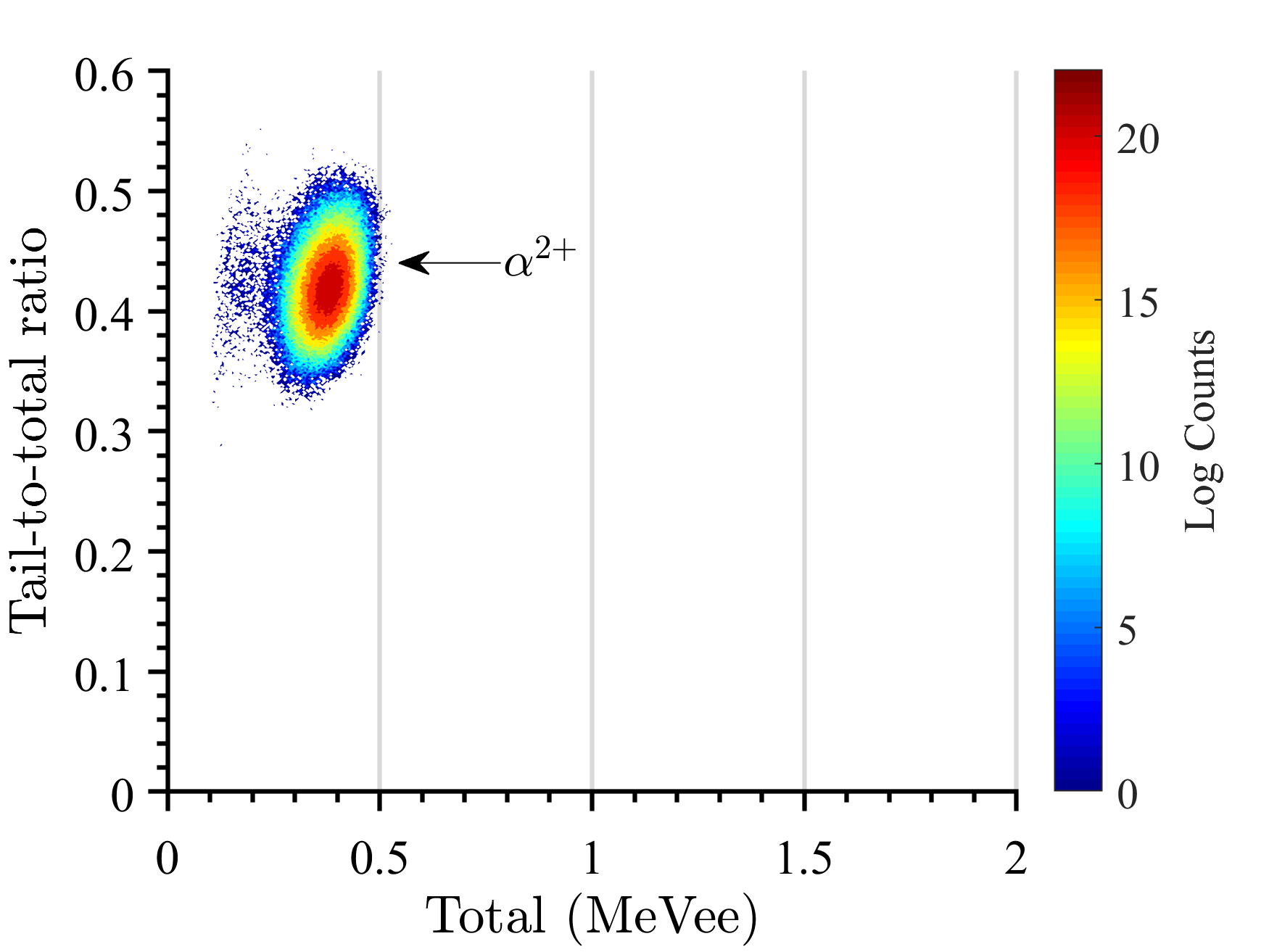}}
		\caption{(a) PSD scatter-density plot of the D-T source. (b) PSD scatter-density plot of the $^{241}$Am alphas. }
		\label{f:DT_Alpha_PSD}
	\end{figure}

The PSD capability of the stilbene-d$_{12}$ is shown in Figure~\ref{f:DT_Alpha_PSD}. The distinct regions in the PSD scatter-density plot of the D-T neutron source (Figure~\ref{f:DT_Alpha_PSD} (a)) showed the capability of the stilbene-d$_{12}$ to distinguish different particles. The uppermost region represents the deuterium-ion pulses from the neutron-deuteron scattering and the lowest region represents the electron pulses from gamma-ray Compton scattering. The small area under the deuterium-ion region is due to the proton pulses via the n(d,2n)p break-up reactions. One may also notice a small hump at around 1~MeVee, which is due to the alpha pulses produced via $^1_0n$($^{12}_6$C,$^{9}_4$Be)$^4_2\alpha$ break-up reactions. Figure~\ref{f:DT_Alpha_PSD} (b) shows the PSD plot of the $^{241}$Am source, the $\alpha$ PSD parameter (TTR) is consistent with the TTR produced by alphas from carbon break-up in D-T PSD scatter-density plot.\par

In Figure~\ref{f:DT_Alpha_PSD}(a), we observe that the alpha pulses from the carbon break-up reactions partially overlap with the deuterium-ion pulses, and they are difficult to distinguish using the PSD charge integration method. However, these overlapped pulses did not affect the calculated scintillation time constants and the neutron light output response. As described in Section 2.1, the deuterium-ion pulses selected for the characterization of the time constants were in the 1.5-6 MeVee range. Therefore, the overlapped alpha pulses (0.5-1 MeVee range) were rejected. As for the light output response measurements, the carbon break-up reactions $((n((_6^12)C,(_4^9)Be )(_2^4)\alpha)$ does not produce any scattered neutrons to be detected by the EJ-309 detectors. Therefore, these alpha pulses will not affect the characterization result of the neutron light output response.

\begin{figure}[!htb]
		\centering
		\subfloat[]{\includegraphics[width=0.45\textwidth]{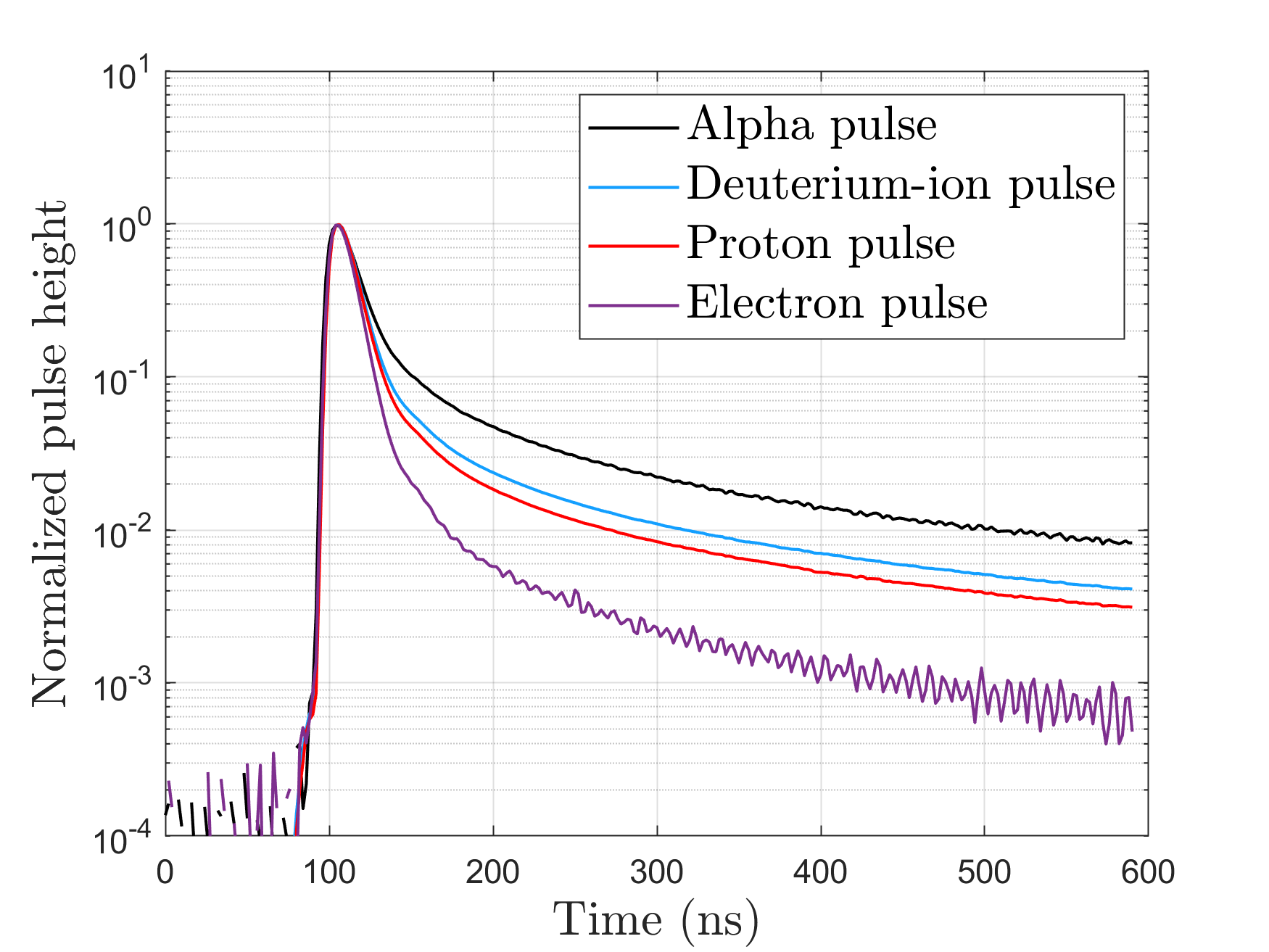}}
		\subfloat[]{\includegraphics[width=0.45\textwidth]{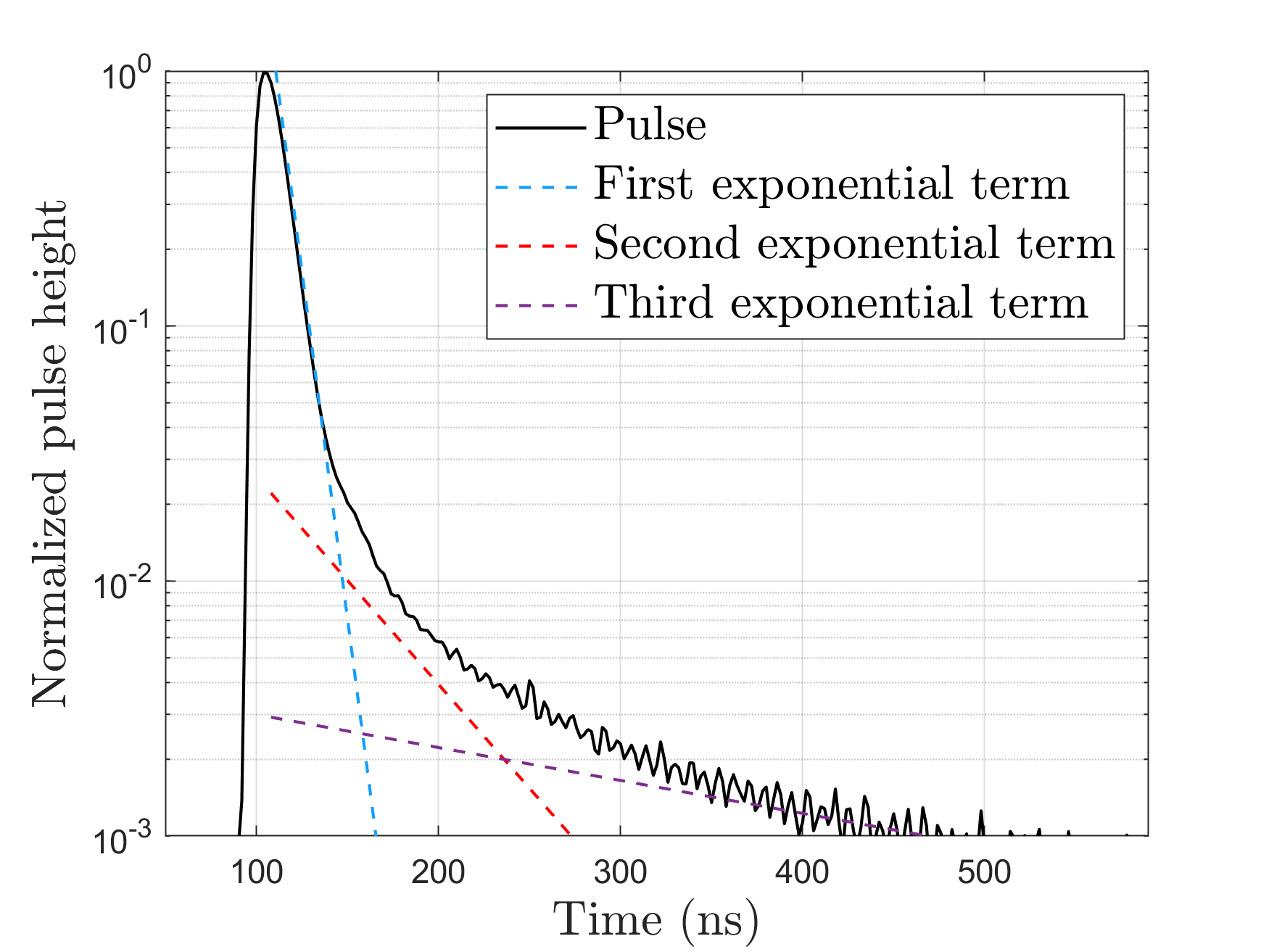}}\\
		\subfloat[]{\includegraphics[width=0.45\textwidth]{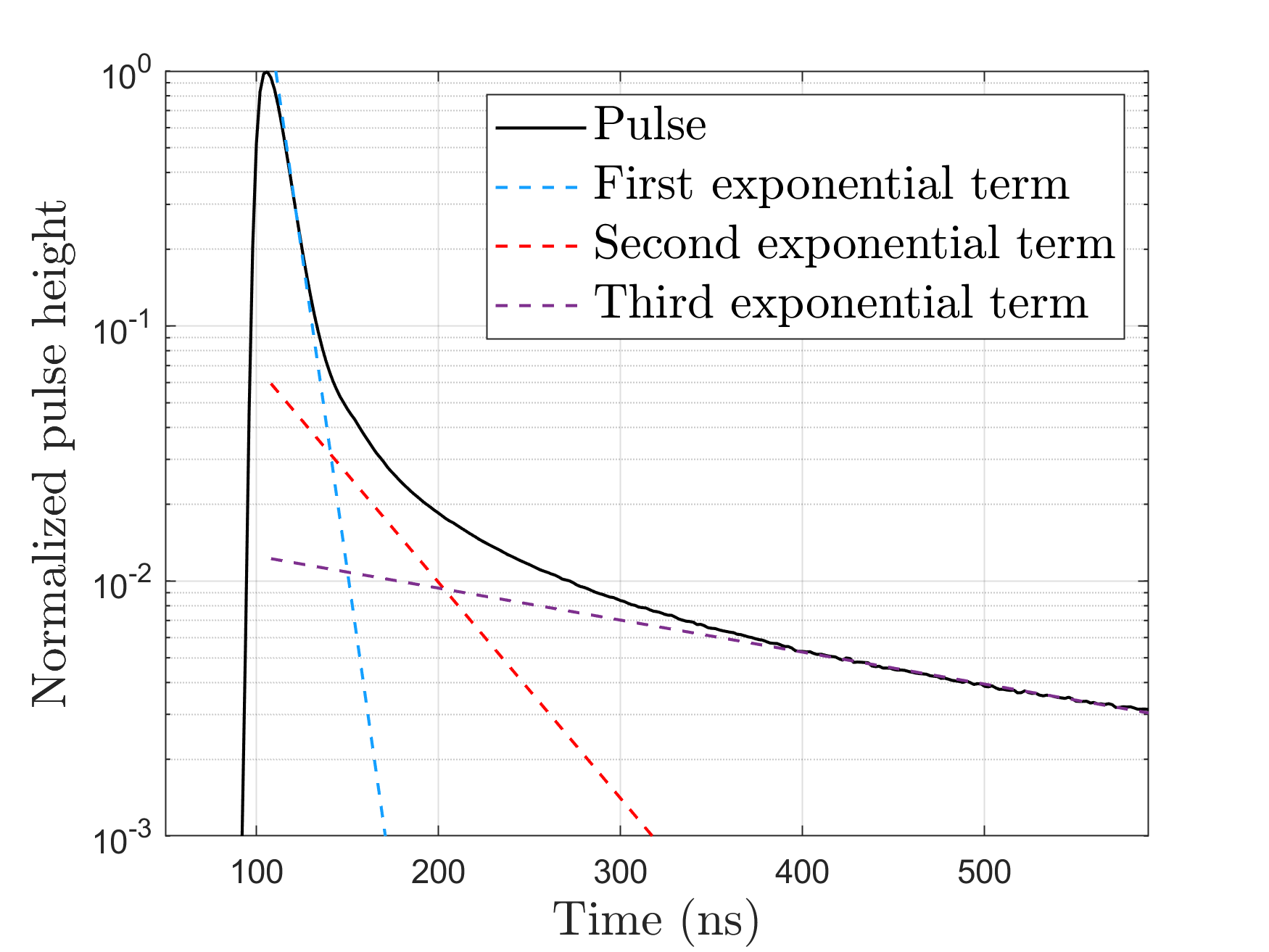}}
		\subfloat[]{\includegraphics[width=0.45\textwidth]{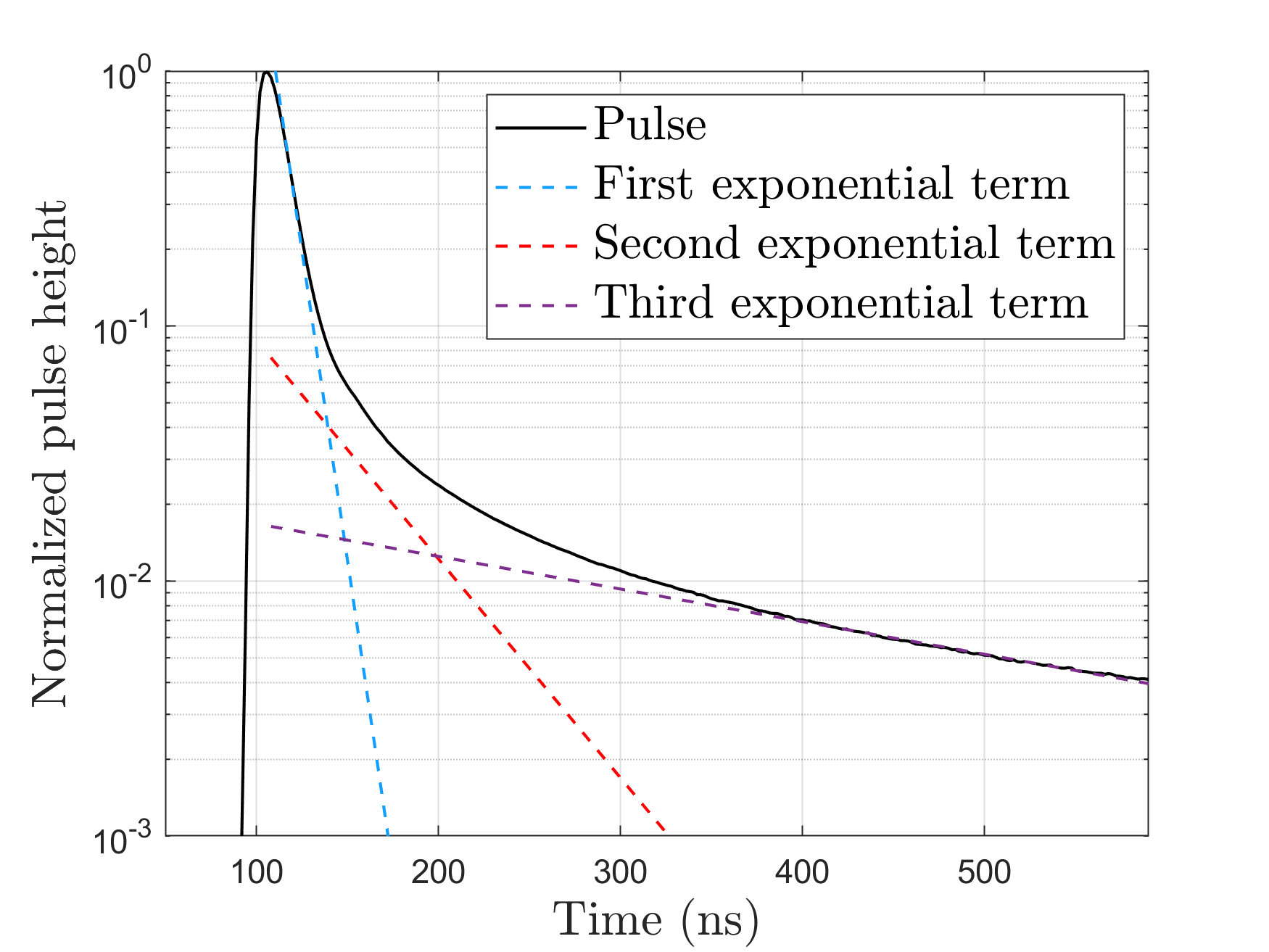}}\\
		\subfloat[]{\includegraphics[width=0.45\textwidth]{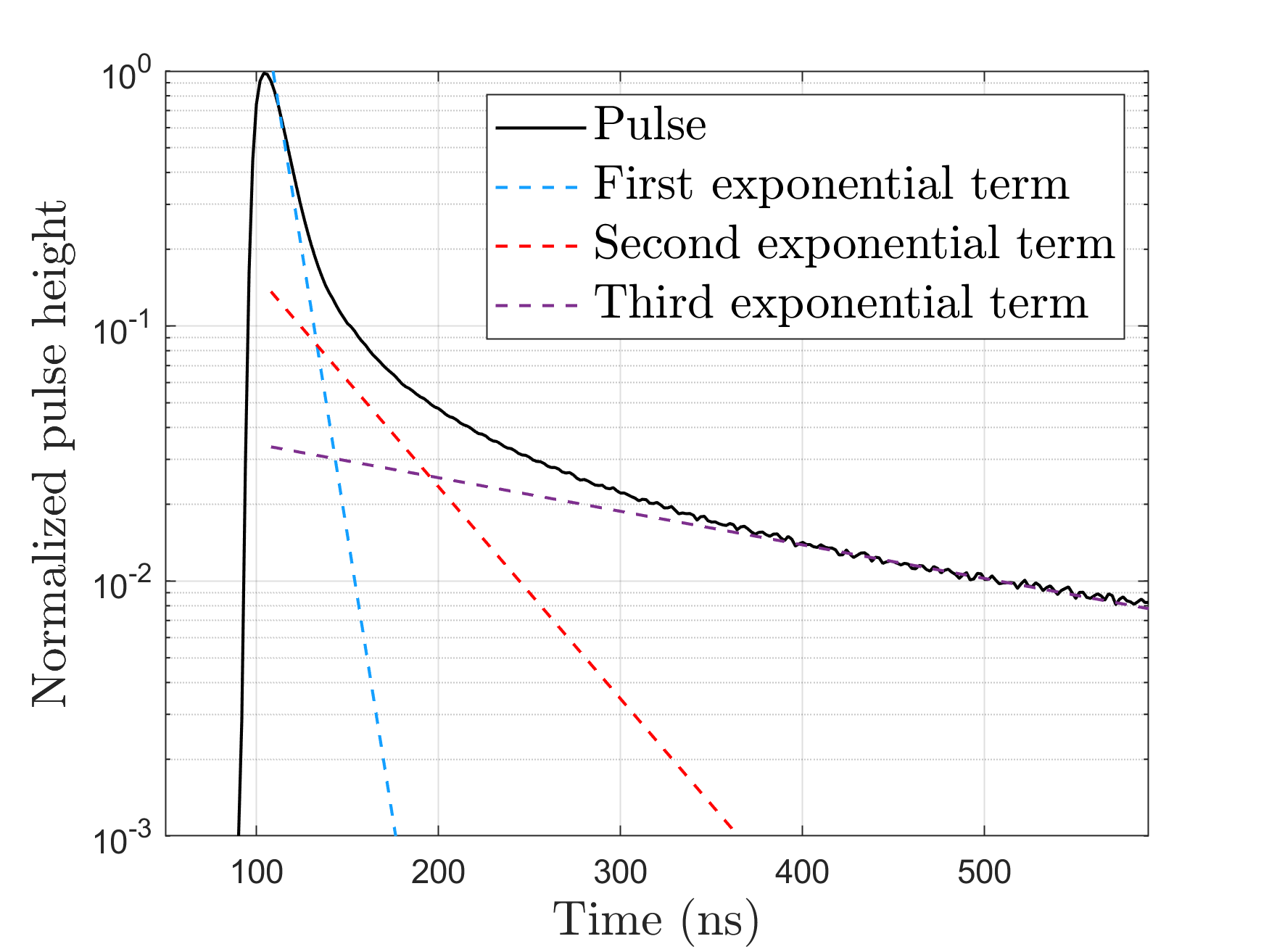}}
		\caption{(a) Averaged pulses with different pulse types. Exponential fit of the (b) electron, (c) proton, (d) deuterium-ion, and (e) alpha pulses.}
		\label{f:Pulse_average_comparison}
	\end{figure}

Using the PSD methods described in the previous sections, we selected electron pulses, proton pulses, and deuterium-ion pulses in the 1.5-6~MeVee range from Figure~\ref{f:DT_Alpha_PSD} (a) to study the quenching effects on the different recoil ions. The three types of pulses are well-separated in this light output range. We also selected alpha pulses from the $^{241}$Am response (Figures~\ref{f:DT_Alpha_PSD} (b)).  We normalized the pulses to the pulse maximum and averaged pulses of the same type to obtain the pulse templates shown in Figure \ref{f:Pulse_average_comparison} (a). We fit the pulse templates using Equation~\ref{eqn: light_exp_decay}. Figures~\ref{f:Pulse_average_comparison} (b) (c) (d) (e) show the light amplitudes and time constants of the fast and slow components of light pulses in response to Compton electrons, protons, deuterium-ions, and alpha particles. The few data-point samples at the rising edges did not allow to obtain an accurate fit of the raising component of the pulse and, therefore, we chose to fit only the three exponential decay terms. 

\begin{table}[htbp!]
    \centering
    \caption{Parameters of the exponential fit for various pulse types. The third exponential term is negligible for electron pulses. }
    \begin{tabular}{ |c|c|c|c|c| }
    \hline 
     & Electron & Proton & Deuterium-ion & Alpha\\
    \hline 
    $\tau_1$ (ns) & $8.11\pm0.10$ & $9.03\pm0.10$ & $9.19\pm0.11$ & $9.87\pm0.11$\\
    $\tau_2$ (ns) & $50.82\pm9.97$ & $50.08\pm3.89$ & $49.78\pm3.05$ & $51.86\pm1.52$\\
    $\tau_3$ (ns) & N/A & $346.5\pm72.9$ & $339.1\pm53.0$ & $330.5\pm22.5$\\
    $q_1$  & $6.5\pm1.2\times10^6$ &  $1.8\pm0.3\times10^6$  &  $1.5\pm0.2\times10^6$ &  $5.9\pm0.8\times10^5$ \\
    $q_2$ & $10.34\pm5.54$ & $26.03\pm5.59$ & $33.36\pm5.66$ & $58.19\pm4.56$\\
    $q_3$ & N/A & $5.81\pm0.98$ & $7.69\pm0.97$ & $15.36\pm0.86$\\
    \hline 
    \end{tabular}
    \label{t:exponential_fit_parameter}
\end{table}

The obtained fitting parameters are reported in Table~\ref{t:exponential_fit_parameter}. We can observe that the time constants ($\tau_1$, $\tau_2$, $\tau_3$) are comparable for different particle types. These results indicate that the ionization quenching does not affect the light decay time. Conversely, the amplitude of the fast and slow components are strongly dependent on the radiation type, as expected. The relative amplitude of the fast component decreases as the LET value of the particles increases.
These results confirm J. B. Birks' observations~\cite{Birks1965}.

\subsection{Neutron light output response}

\begin{figure}[!htb]
		\centering
		\subfloat[]{\includegraphics[width=0.55\textwidth]{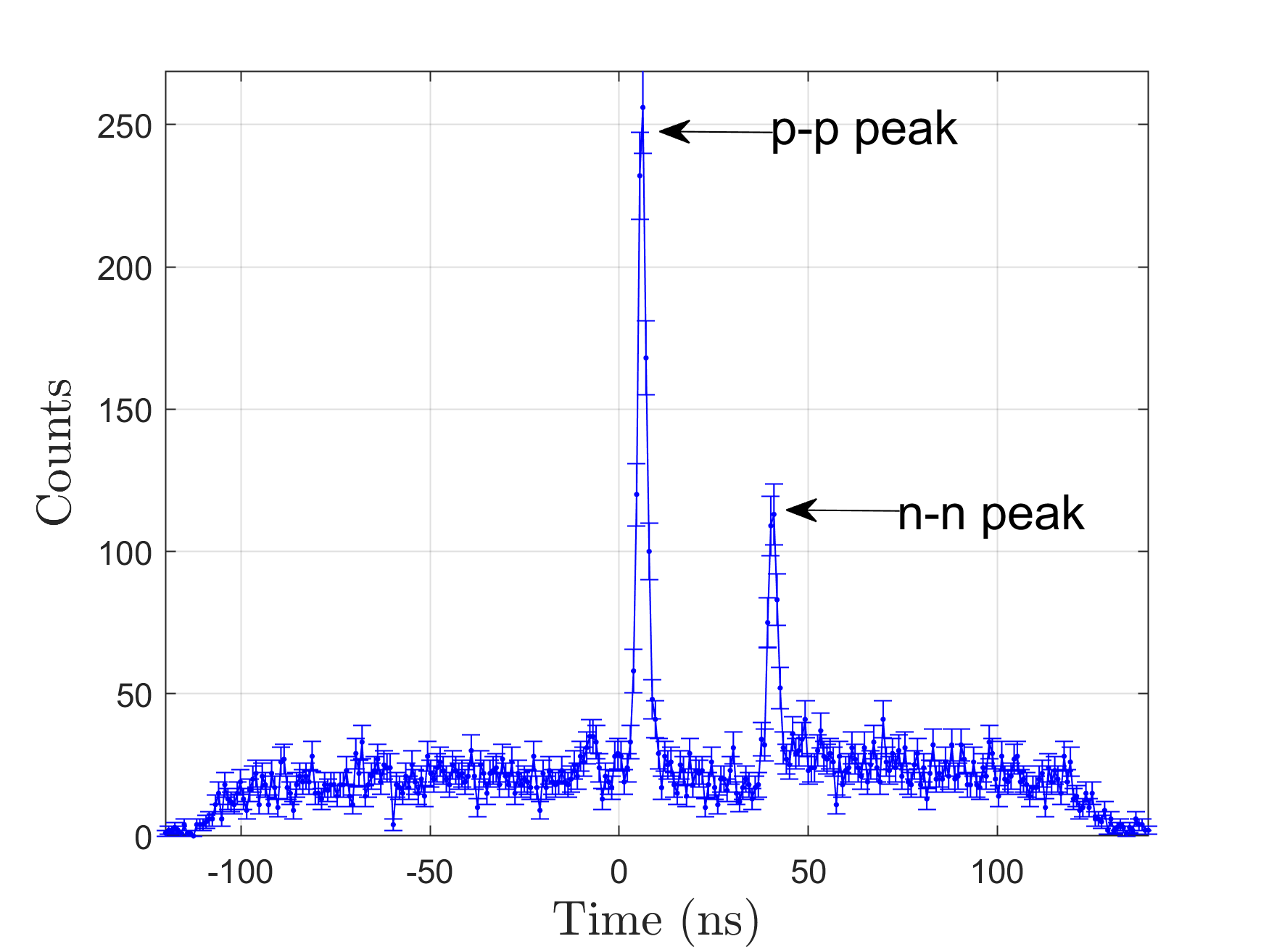}}
		\subfloat[]{\includegraphics[width=0.55\textwidth]{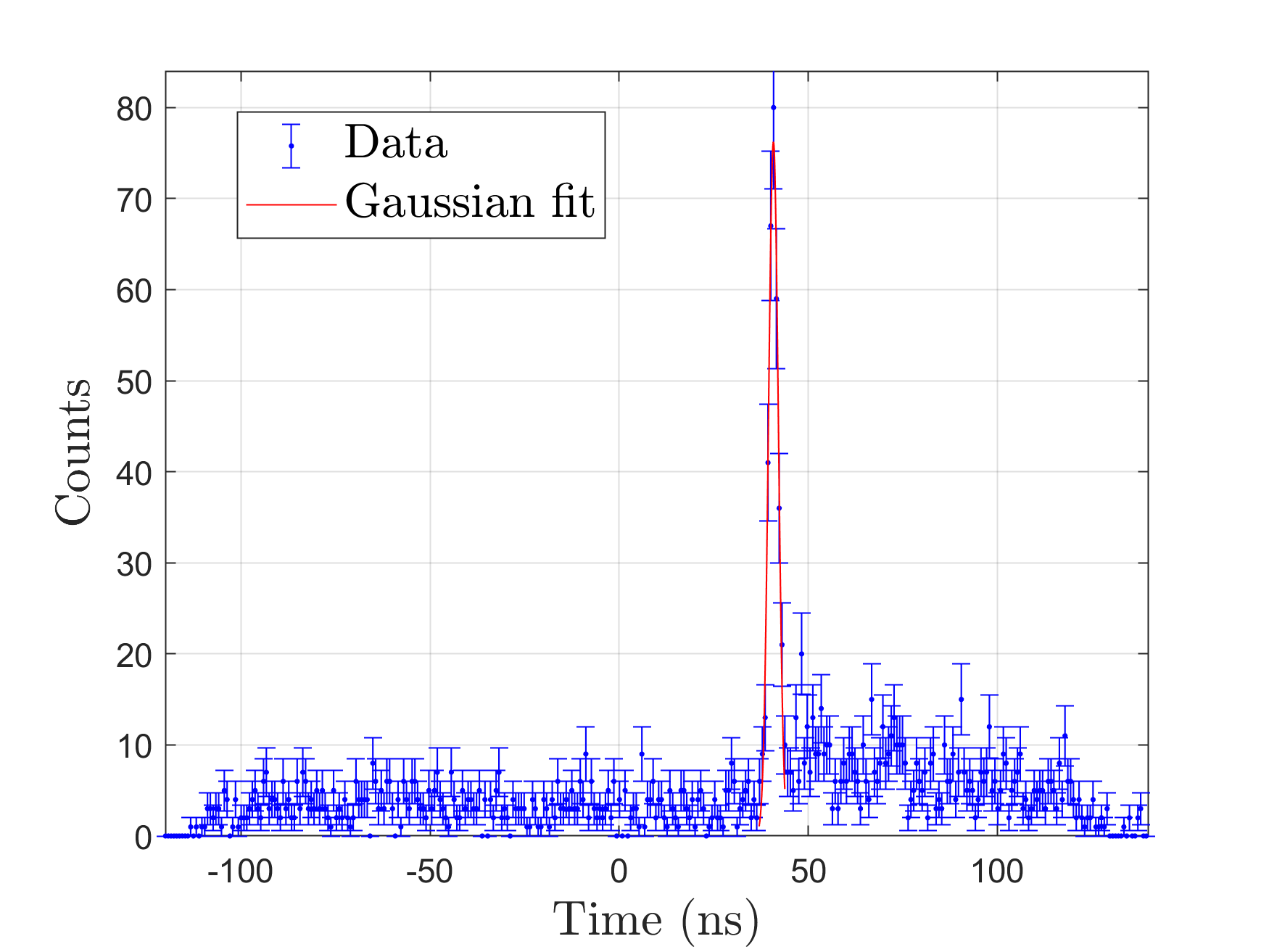}}
		\caption{(a) Coincidence TOF spectrum for the 30 degree measurement. (b)  Coincidence TOF spectrum for the 30 degree measurement after gamma rejections with PSD. }
		\label{f:30degree_tof_spec}
	\end{figure}

Figure~\ref{f:30degree_tof_spec} (a) shows the TOF spectrum for the 30\degree scattering angle measurement, with a 2~m flight path. The gamma-ray would take 6.7 ns to reach the EJ-309 detectors after interacting with the stilbene-d$_{12}$. Therefore, the peak near the origin is due to the gamma-gamma coincidence events. The neutrons scattered by 30\degree retain 12.31 MeV from the primary neutron, hence their flight time is 41.2 ns, which agrees with the peak at approximately 40 ns in Figure~\ref{f:30degree_tof_spec} (a) . 

\begin{figure}[htb!]
    \centering
    \includegraphics[width=0.55\textwidth]{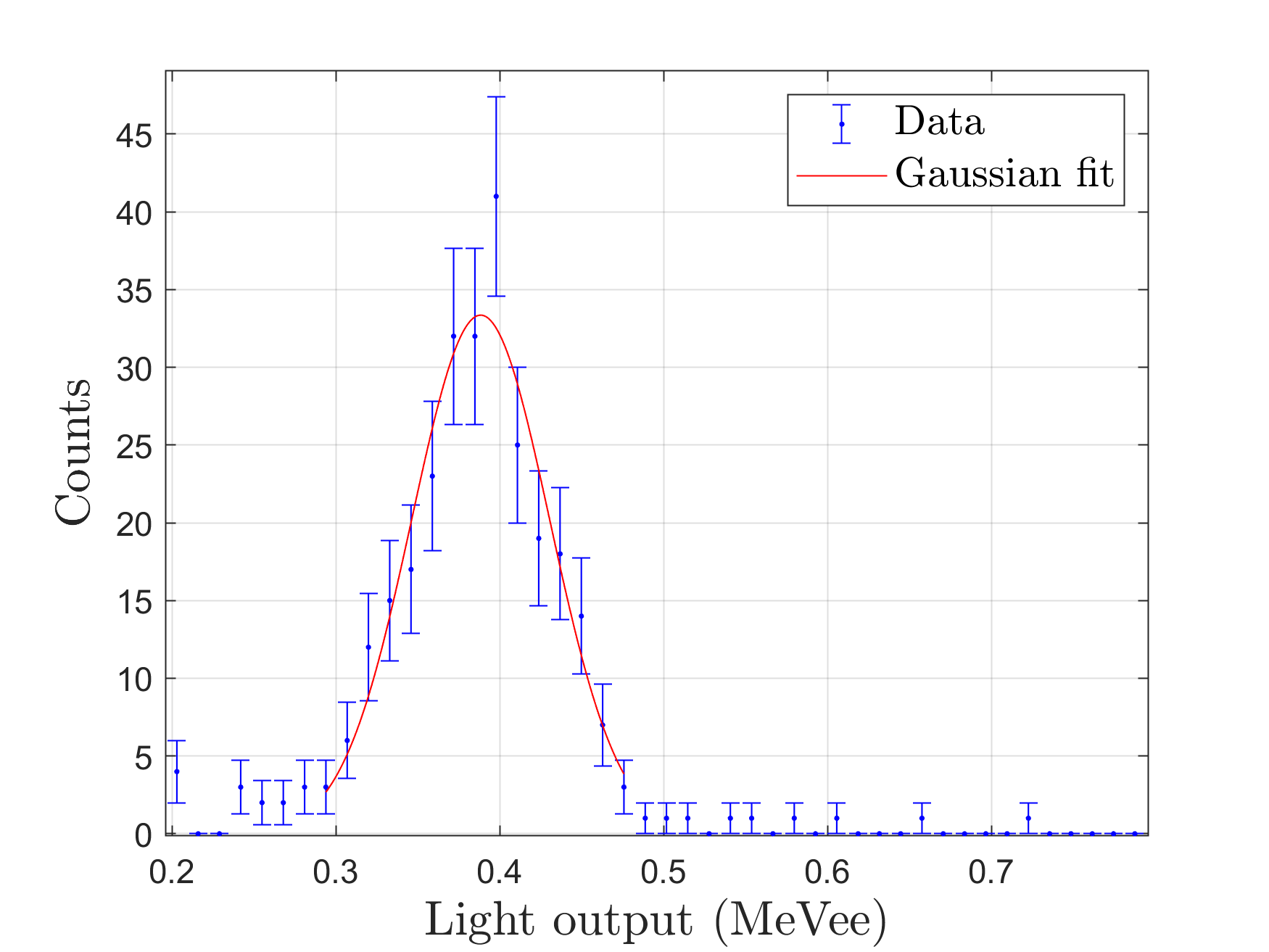}
    \caption{Stilbene-d$_{12}$ light output spectrum for pulses under the neutron-neutron coincidence TOF peak}
    \label{f:30degree_lightoutput}
\end{figure}

\begin{figure}[!htb]
		\centering
		\subfloat[]{\includegraphics[width=0.45\textwidth]{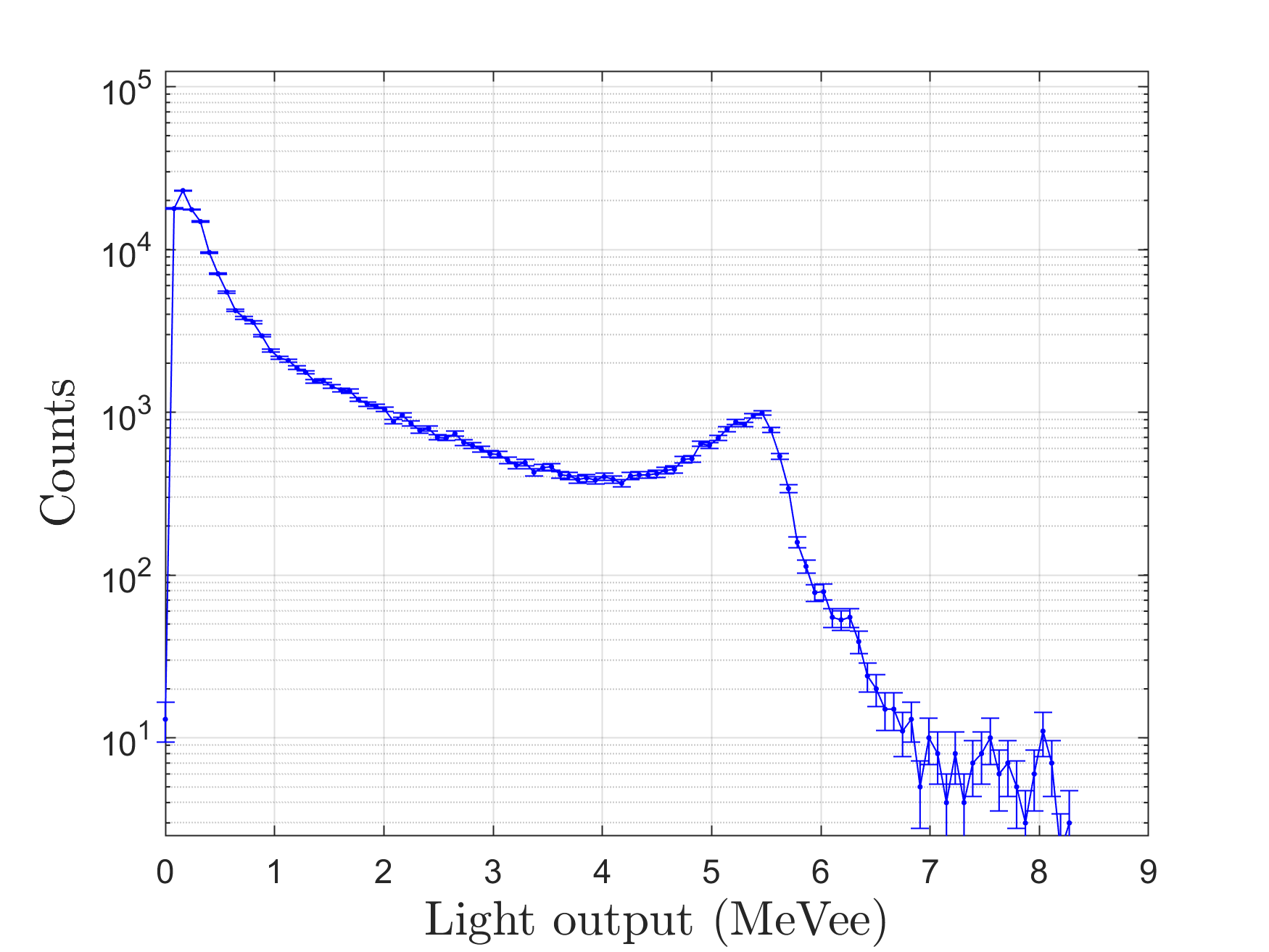}}
		\subfloat[]{\includegraphics[width=0.45\textwidth]{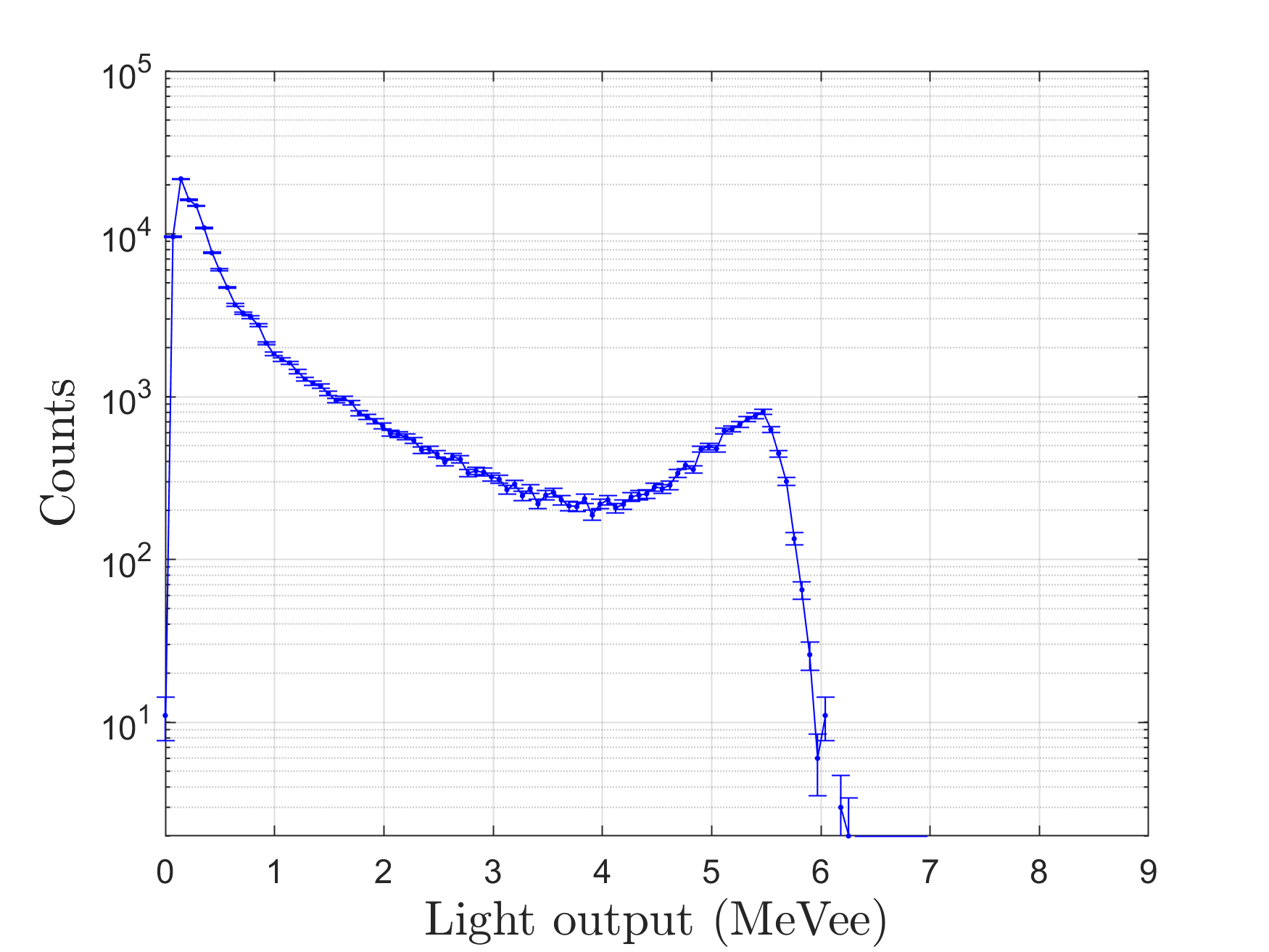}}
		\caption{(a) Light output spectrum for the D-T source after rejecting gamma ray pulses. (b) Light output spectrum for the D-T source after rejecting gamma ray and proton pulses. }
    \label{f:DT_pid}
	\end{figure}

In Figure \ref{f:30degree_tof_spec} (b), we performed the PSD on both stilbene-d$_{12}$ and EJ-309 detectors to reject gamma-ray pulses. Therefore, we can only observe the neutron-neutron peak. Neutrons transfer 1.79 MeV to deuterium-ions after the 30\degree scattering. Figure~\ref{f:30degree_lightoutput} shows the light output spectrum produced by all coincidence events under the PSD-selected neutron-neutron peak
corresponding to the 1.79 MeV deuterium-ion. Table~\ref{t:neutron_lightouput_data} shows the light output values measured at the other angles, using a similar procedure. We also performed a direct measurement of D-T generated 14.1~MeV neutrons, which corresponds to a maximum energy deposited by deuterium ion of 12.53~MeV ($E_r$ in Equation~\ref{eqn: scattering_kine2} when $\theta$ equals 180$^o$ and A equals to 2). Figure~\ref{f:DT_pid} shows the light output spectrum in response to 14.1 MeV neutrons after discriminating gamma rays through the quadratic polynomial PSD discrimination line. Figure~\ref{f:DT_pid} (a) includes both scattered deuterium ions and protons produced by the neutron-deuteron beak-up reactions. These proton pulses are less quenched than the deuterium-ion ones and are hence visible on the right-left side of the light-output spectrum. Figure~\ref{f:DT_pid} shows the response to deuterium-ions only, obtained after rejecting the proton pulses through PSD. 

\begin{table}[h!]
    \centering
    \caption{Deuterium-ion energies and measured light output values for six measurement angles.}
    \begin{tabular}{ |c|c|c| }
    \hline 
    Scattering angle  & Deuterium-ion energy & Measured light output value\\
    (\degree) & (MeV) & (MeVee)\\
    \hline 
    15 & 0.47$\pm$0.09 & 0.10$\pm$0.05\\
    30 & 1.79$\pm$0.16 & 0.37$\pm$0.04\\
    50 & 4.38$\pm$0.21 & 1.33$\pm$0.08\\
    65 & 6.19$\pm$0.10 & 2.11$\pm$0.12\\
    120 & 11.44$\pm$0.03 & 4.67$\pm$0.14\\
    140 & 12.11$\pm$0.02 & 5.14$\pm$0.15\\
    N/A & 12.53 & 5.46 \\
    \hline 
    \end{tabular}
    \label{t:neutron_lightouput_data}
\end{table}

\subsection{Alpha Light Output Response}

\begin{figure}[!htbp]
		\centering
		\subfloat[]{\includegraphics[width=0.33\textwidth]{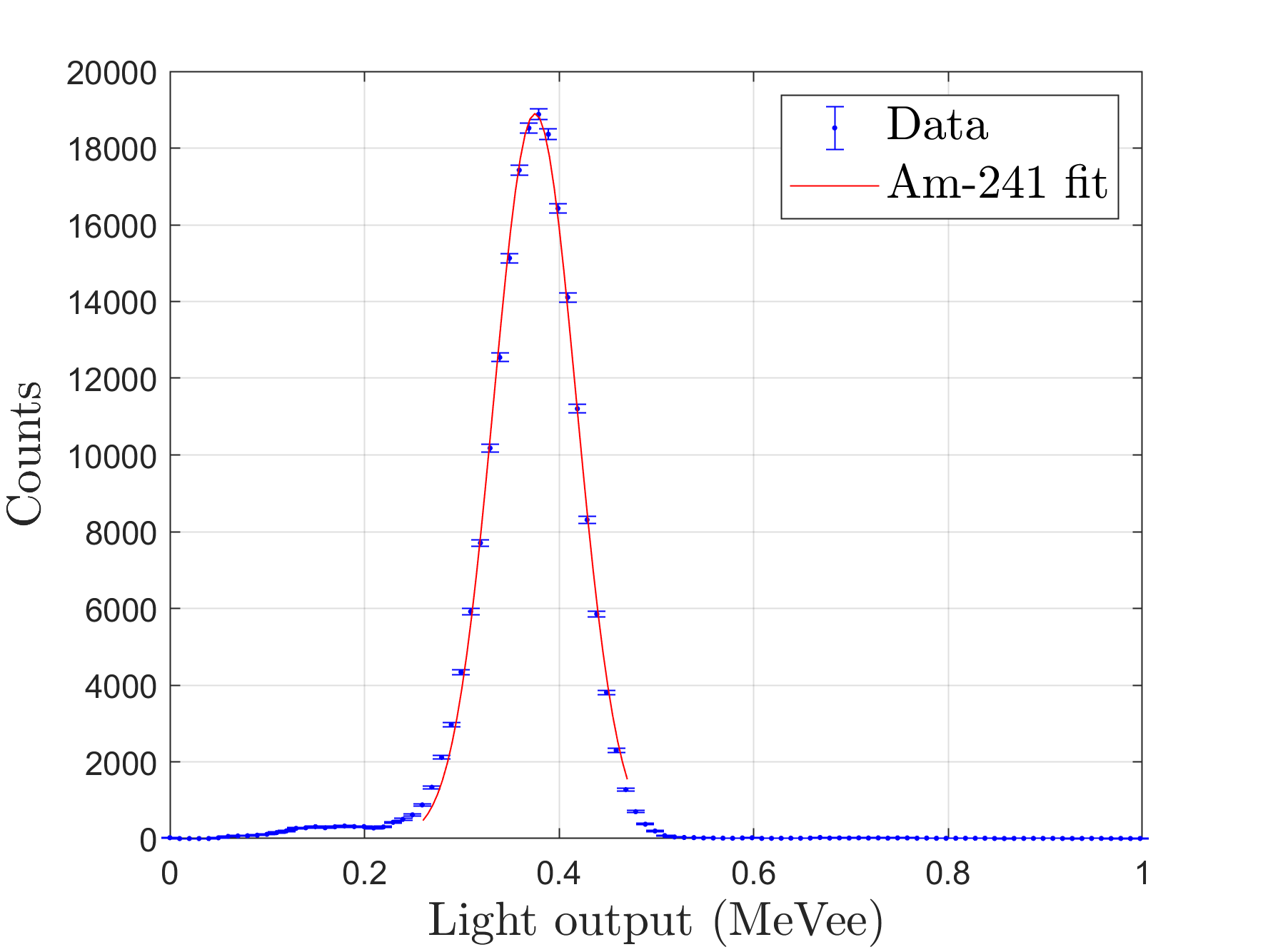}}
		\subfloat[]{\includegraphics[width=0.33\textwidth]{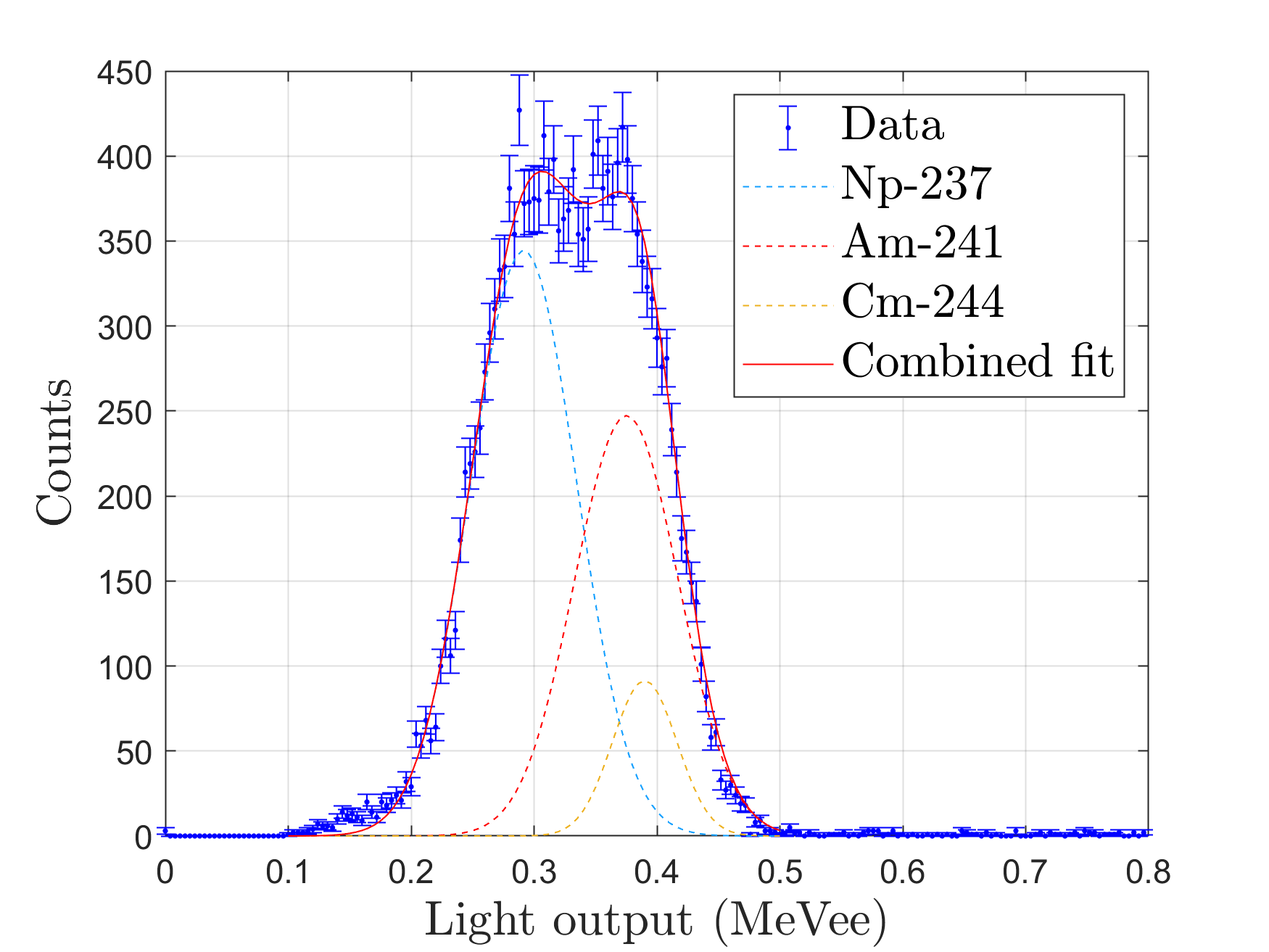}}
		\subfloat[]{\includegraphics[width=0.33\textwidth]{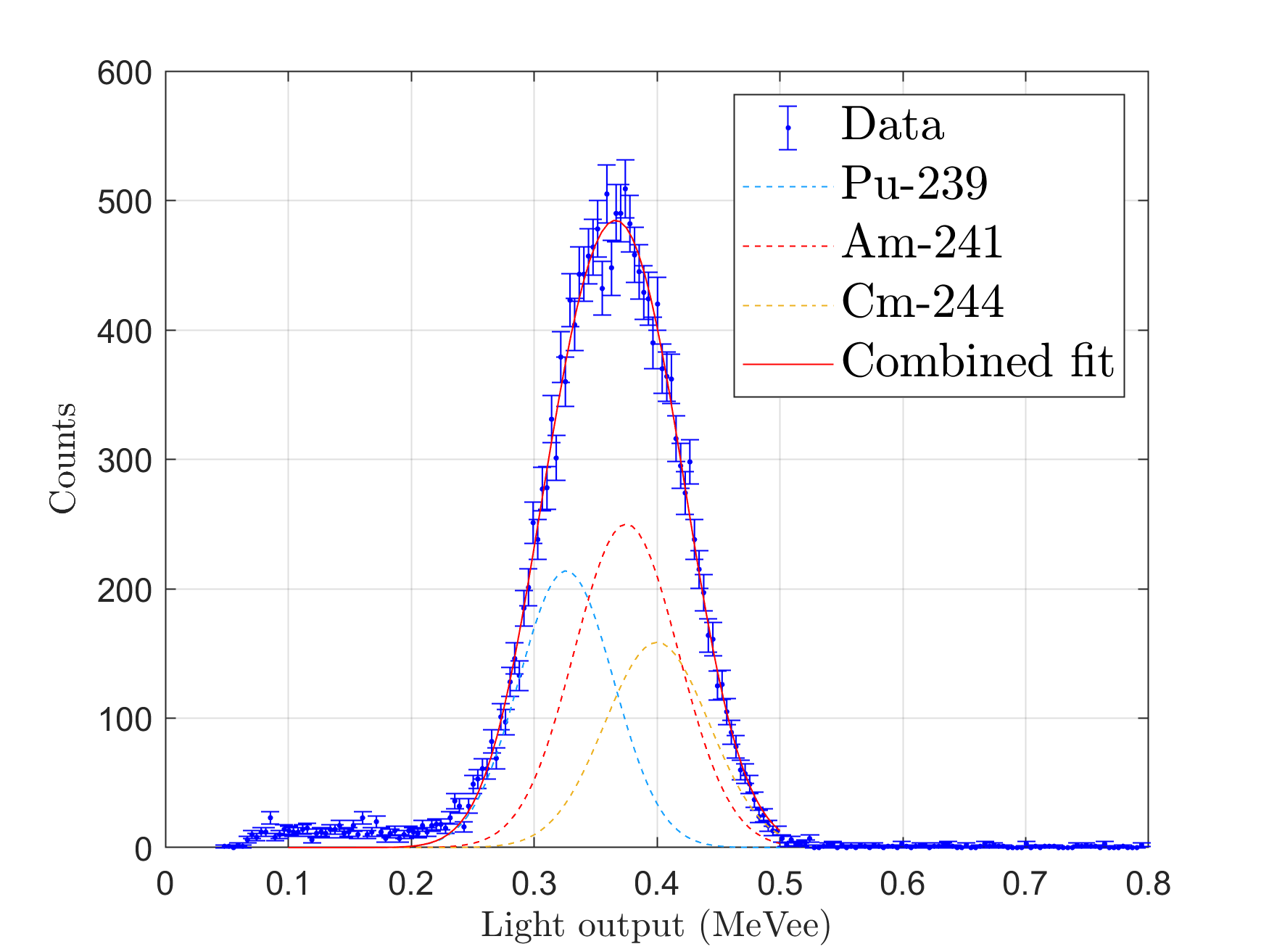}}\\
		\caption{(a) Light output spectrum of the $^{241}$Am source. (b) Light output spectrum of the $^{237}$Np, $^{241}$Am, and $^{244}$Cm mixed source. (c) Light output spectrum of the $^{239}$Pu, $^{241}$Am, and $^{244}$Cm mixed source. }
		\label{f:alpha_spectrum}
	\end{figure}

Figure~\ref{f:alpha_spectrum} shows the measured light output spectra in response to three alpha sources. The energy resolution of the stilbene-d$_{12}$ crystal in response to monoenergetic alpha particles, calculated as $\frac{\delta E}{E}$ (${\delta E}$ full width at half maximum), is approximately 26\% (Figure~\ref{f:alpha_spectrum} (a)). With such energy resolution, the discrete alpha energies of the two mixed alpha sources cannot be separated. Therefore, we fit the measured spectrum with the linear combination of three Gaussian distributions, as shown in Figure \ref{f:alpha_spectrum} (b) and (c), and acquired the light output values for discrete alpha energies corresponding to the peaks of the Gaussian distributions. The light output values in response to alpha particles in the 5-6 MeV energy range are reported in Table~\ref{t:alpha_lightouput_data}.

\begin{table}[h!]
    \centering
    \caption{Alpha energies and the corresponding light output values.}
    \begin{tabular}{ |c|c|c| }
    \hline 
    Alpha source  &  $\alpha$ Energy & Measured light output value\\
     & (MeV) & (keVee)\\
    \hline 
    Np-237 & 4.959 & 291.5$\pm$43.0\\
    Pu-239 & 5.244 & 326.2$\pm$38.5\\
    Am-241 & 5.637 & 375.0$\pm$42.3\\
    Cm-244 & 5.902 & 400.0$\pm$42.4\\
    \hline 
    \end{tabular}
    \label{t:alpha_lightouput_data}
\end{table}

\subsection{Ionization Quenching Parameters}

\begin{figure}[htbp!]
    \centering
    \includegraphics[width=0.75\textwidth]{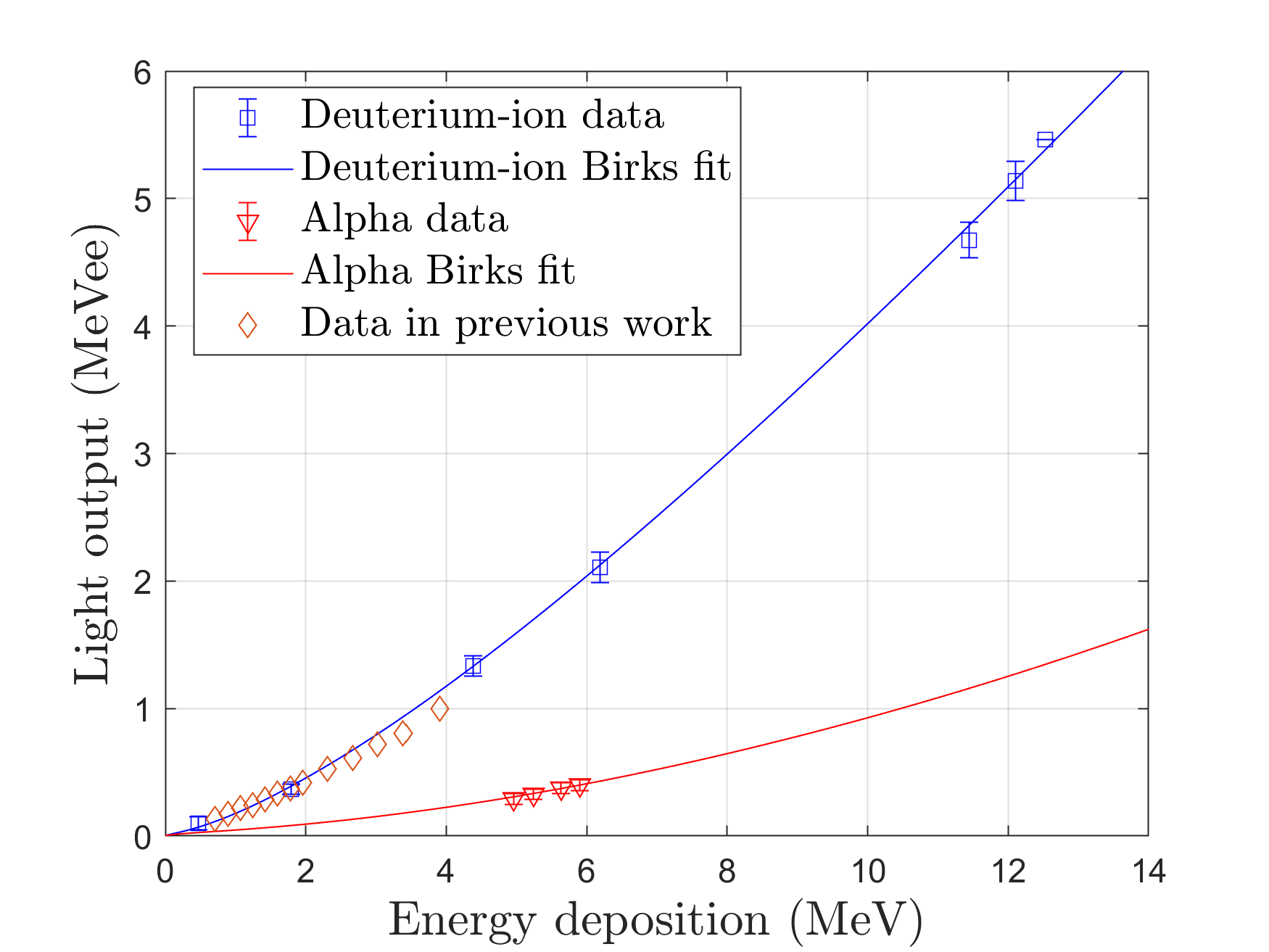}
    \caption{Stilbene-d$_{12}$ light output responses to deuterium-ions and alpha particles.}
    \label{f:d_a_lightoutput_response}
\end{figure}

We fit the measured light output response to neutron-scattered deuteron-ion recoils and and alpha particles using the Birks' model (Equation~\ref{eqn:Birks_model}) to study the stilbene-d$_{12}$ quenching to different particles.
Table~\ref{t:d_a_Birks_coeff} summarizes the measured light output response and Figure~\ref{f:d_a_lightoutput_response} shows measured data and the light output curve obtained fitting the Birks' model to the measured data. The extended neutron light output response agrees well with the response of the 32~cm$^3$ crystal~\cite{Gaughan2021} measured up to 4.4 MeV neutron energy. 

\begin{table}[h!]
    \centering
    \caption{Coefficients of the Birks' model and their relative uncertainties for the neutron and alpha light output response. }
    \begin{tabular}{ |c|c|c| }
    \hline 
     & Coefficient & Value \\
    \hline 
    Deuterium-ion & A (MeVee MeV$^{-1}$) & $0.75\pm 0.05$ \\
    & B (mg MeV$^{-1}$ cm$^{-2}$) & $5.80\pm 0.10$ \\
    \hline 
    Alpha & A (MeVee MeV$^{-1}$) & $4.00\pm 0.05$\\
    & B (mg MeV$^{-1}$ cm$^{-2}$) & $49.60\pm 0.10$ \\
    \hline 
    \end{tabular}
    \label{t:d_a_Birks_coeff}
\end{table}

\section{Conclusions}\label{s:conclusion}

Stilbene-d$_{12}$ is an organic scintillating crystal that exhibits excellent PSD and can perform neutron spectroscopy without time-of-flight. In this work, we characterized the neutron light output response up to 14.1 MeV neutron energy for a recently grown stilbene-d$_{12}$ detector cell of 140 $cm^3$ volume. The neutron response function agrees well with previous data, measured up to 4.4 MeV neutron energy for a smaller crystal. Knowledge of the accurate stilbene-d$_{12}$ light output response to neutrons will allow to derive via simulation its response matrix, which is needed to perform spectrum unfolding.  
We also characterized the stilbene-d$_{12}$ response to alpha particles measuring the light output corresponding to 5.637 MeV, 4.959 MeV, 5.902 MeV, and 5.244 MeV alphas emitted by $^{241}Am$, $^{237}Np$, $^{244}Cm$, and $^{239}Pu$, respectively. By fitting the light-output responses to the Birks' model, we found quenching parameters of 5.8$\pm$0.1 mg~MeV$^{-1}$~cm$^{-2}$ and 49.6$\pm$0.1 mg~MeV$^{-1}$~cm$^{-2}$ for deuterium ions, produced by neutron interactions, and alpha particles, as expected due the the higher ionization density of the alpha particles (5-6 MeV) compared to the deuterium ions (1-13 MeV). \par
Highly-quenched ionizing particles are known to have a prominent contribution of the slow fluorescence component, compared to the fast one. This phenomenon was confirmed by measuring and comparing the time constants of the fast and slow fluorescence components of electrons, protons, deuterons, and alpha particles. The decay constants of the two identified slow components are approximately 50 ns and 330 ns, corresponding to the annihilation of excited triplet states and inter-system crossing, respectively, regardless of the interacting particle. The relative intensity of the slow component to the overall light pulse increases with the ionization density because of the quenching of the fast component of the scintillation light.\par
The PSD result of the stilbene-d$_{12}$ in response to 14.1 MeV neutrons demonstrated that the charge integration method cannot distinguish the neutrons and alpha particles well. We are developing a machine-learning (ML) based PSD technique for the stilbene-d$_{12}$ to overcome this issue. The charge integration method uses the integral of different portions of pulses for PSD and may not fully exploit the shape differences of various radiation particles. Conversely, the ML-based method is able to perform the classification of the pulses based on all the time constants featured by the materials. Therefore, an advanced ML algorithm could improve the current particle classification performance of stilbene-d$_{12}$.

\section{Acknowledgements}

This work was funded in part by the Nuclear Regulatory Commission (NRC) Faculty Development Grant 31310019M0011, and NRC fellowship grants NRC-HQ841560020 and NRC-31310018M0029. Crystal growth at LLNL was supported by the United States National Nuclear Security Administration’s Office of Defense Nuclear Nonproliferation Research $\&$ Development. This material is based upon work supported by the U.S. Department of Energy, Office of Science, Office of Nuclear Physics, under Award Number DE-AC05-00OR22725

\medskip

\bibliography{manuscript}

\end{document}